\documentclass[journal,twoside,web]{ieeecolor}

\usepackage{xcolor}
\usepackage[breaklinks=true,colorlinks,bookmarks=false]{hyperref}
\usepackage{dblfloatfix}
\usepackage[caption = false]{subfig}
\usepackage{multirow}
\usepackage{amsmath,amssymb,amsfonts}
\usepackage{graphicx}
\usepackage[ruled,vlined,linesnumbered,resetcount]{algorithm2e}

\usepackage{generic}

\graphicspath{{Figures/}}

\makeatletter
\newcommand{\nosemic}{\renewcommand{\@endalgocfline}{\relax}}
\newcommand{\dosemic}{\renewcommand{\@endalgocfline}{\algocf@endline}}
\newcommand{\pushline}{\Indm\nosemic}
\newcommand{\popline}{\Indp\dosemic}
\let\oldnl\nl
\newcommand{\nonl}{\renewcommand{\nl}{\let\nl\oldnl}}
\makeatother

\date{}

\begin{document}
\title{A multi-stream convolutional neural network for classification of progressive MCI in Alzheimer's disease using structural MRI images}
\author{Mona~Ashtari-Majlan, Abbas Seifi, Mohammad Mahdi~Dehshibi
\thanks{M. Ashtri-Majlan is with the Department of Computer Science, Universitat Oberta de Catalunya, Barcelona, Spain (e-mail: ashtari.mona@gmail.com).}
\thanks{A. Seifi is with the Department of Industrial Engineering, Amirkabir University of Technology, Tehran, Iran (e-mail: aseifi@aut.ac.ir).}
\thanks{M. M. Dehshibi is with the Department of Computer Science, Universitat Oberta de Catalunya, Barcelona, Spain (e-mail: mohammad.dehshibi@yahoo.com).}}

\maketitle

\begin{abstract}
Early diagnosis of Alzheimer’s disease and its prodromal stage, also known as mild cognitive impairment (MCI), is critical since some patients with progressive MCI will develop the disease. We propose a multi-stream deep convolutional neural network fed with patch-based imaging data to classify stable MCI and progressive MCI. First, we compare MRI images of Alzheimer’s disease with cognitively normal subjects to identify distinct anatomical landmarks using a multivariate statistical test. These landmarks are then used to extract patches that are fed into the proposed multi-stream convolutional neural network to classify MRI images. Next, we train the architecture in a separate scenario using samples from Alzheimer’s disease images, which are anatomically similar to the progressive MCI ones and cognitively normal images to compensate for the lack of progressive MCI training data. Finally, we transfer the trained model weights to the proposed architecture in order to fine-tune the model using progressive MCI and stable MCI data. Experimental results on the ADNI-1 dataset indicate that our method outperforms existing methods for MCI classification, with an F1-score of 85.96\%.
\end{abstract}


\section{Introduction}
\IEEEPARstart{A}{lzheimer's} disease (AD) is a progressive neurodegenerative disorder that is one of the leading causes of dementia in the elderly. According to~\cite{barnes2011projected}, this disorder affects over 30 million people worldwide. Early diagnosis of this disease and its prodromal stage, also known as mild cognitive impairment (MCI), is crucial since 10\% to 15\% of MCI patients progress to AD, which is classified as progressive MCI (pMCI)~\cite{petersen2009mild}.

\begin{figure}[!htb]
    \centering
    \subfloat[]{\includegraphics[width=0.40\linewidth]{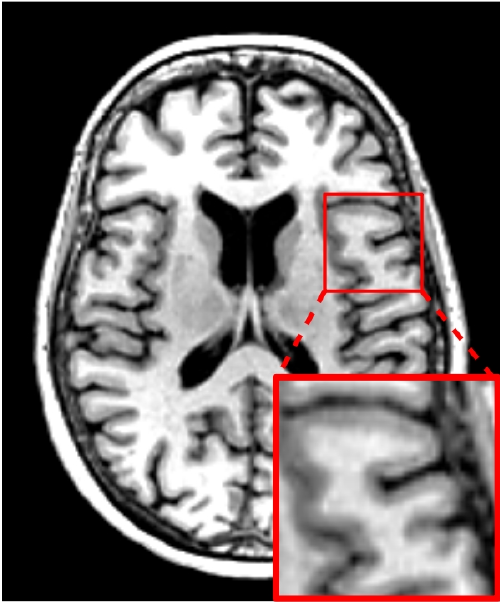}\label{fig:CN}}\qquad
    \subfloat[]{\includegraphics[width=0.382\linewidth]{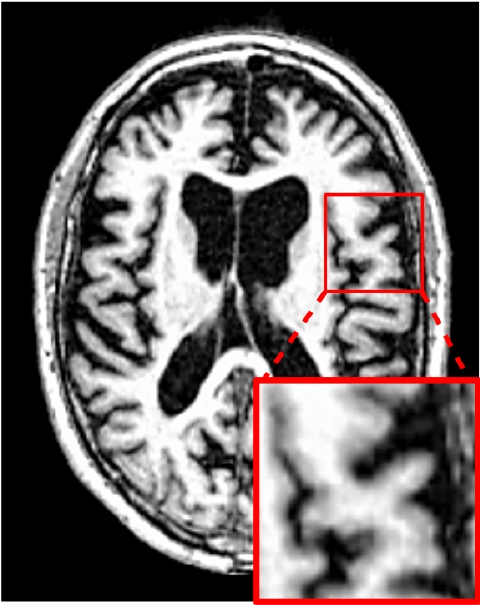}\label{fig:AD}}
    \caption{MRI samples of (a) Cognitively Normal (CN) and (b) AD classes from Alzheimer's Disease Neuroimaging Initiative database (ADNI-1)~\cite{jack2008alzheimer}. In order to demonstrate the subtle brain atrophy, we highlighted the affected regions.}
    \label{fig:MRIs}
\end{figure}

As AD advances, several brain regions develop structural deformation and atrophy~\cite{deture2019neuropathological}. Structural Magnetic Resonance Imaging (sMRI) is one of the most widely employed neuroimaging tools for predicting this disorder through identifying brain atrophy~\cite{hett2021multi} (see Fig.~\ref{fig:MRIs}). In addition to sMRI (referred to as MRI in this paper), non-invasive biomarkers such as (1) demographic information (\textit{e.g.}, age and education)~\cite{liu2019joint}, and (2) cognitive test scores~\cite{hett2021multi} can also be used to provide possible discriminative information for diagnosing AD in the early stages. Several studies~\cite{hett2021multi,lian2020hierarchical,li2019novel,liu2020multi,qiu2020development} have addressed the MCI-to-AD conversion issue using neuroimaging methods in conjunction with the biomarkers.

Conventional methods in medical image processing typically use prior knowledge to segment brain images into Regions of Interest (ROI)~\cite{cabral2015predicting} or Voxel-Based Morphometry (VBM)~\cite{ashburner2000voxel} to predict AD. While these methods can classify stable MCI (sMCI) and pMCI, the focus was mainly on Alzheimer's disease. Deep learning algorithms, on the other hand, have made it possible for researchers to combine feature extraction, dimensionality reduction, and classification in an end-to-end way. These algorithms also outperform conventional methods in identifying AD patterns because they can discover hidden representations among multiple regions of neuroimages. Hence, these models have gained prominence in analysing Alzheimer's disease.

In this paper, we propose a multi-stream convolutional neural network (CNN) for classifying sMCI and pMCI, which is fed with patch-based imaging data extracted using a novel data-driven technique. To accomplish so, we first use the multivariate T2 Hotelling test to compare MRI images of AD and cognitively normal (CN) individuals in order to identify distinct anatomical landmarks. Following that, the statistical test is performed on textural and statistical features extracted from MRI images. These landmarks are then used to generate $19\times19\times19$ patches, with each landmark serving as the centre of the patch. Finally, the extracted patches are fed into the proposed multi-stream CNN to classify MRI images. To compensate for the lack of pMCI training data, we first train the proposed architecture using AD/CN images that are anatomically similar to the pMCI ones. Then, we transfer the weights of the trained model to the proposed architecture in order to fine-tune the model using pMCI and sMCI data. The contribution of this study is two-fold:
\begin{enumerate}
    \item Rather than utilising non-rigid registration to identify anatomical landmarks in the brain, we propose employing rigid registration to reduce computational complexity and eliminate morphological structure deformations that could cause inherent errors in the classification step. Therefore, we partition each MRI image during the anatomical landmark detection phase and perform the T2 Hotelling test on these partitions to capture subtle anatomical differences, to account for more information, and to reduce the impact of potential errors caused by inter-subject brain shape variations. Finally, the identified statistically significant landmarks centres from the partitions are used as anchoring forces for selecting the patches fed into the proposed multi-stream CNN.
    \item We propose using transfer learning to classify pMCI/sMCI classes in training the proposed architecture to overcome the complexity of learning caused by the subtle structural changes in MCI brains compared to AD and CN. In the Ablation study, we also demonstrate the importance of employing transfer learning. Furthermore, we address intrinsic errors caused by inter-subject brain shape differences by conducting experiments to determine an ideal image patch size in order to feed to the proposed CNN model.
\end{enumerate}

The rest of this paper is organised as follows: Section~\ref{sec:survey} surveys the previous studies. Section~\ref{sec:method} describes the proposed method. Experimental results are given in Section~\ref{sec:experiment}. Finally, the paper is concluded in Section~\ref{sec:conclusion}.

\section{Literature review}\label{sec:survey}
Mild cognitive impairment (MCI) is a stage of cognitive decline that occurs between the predicted cognitive loss associated with normal ageing and the more severe decline associated with dementia. MCI may result in a higher level of developing dementia caused by Alzheimer’s disease if the anatomical changes in the brain are proactive. Progressive MCI differs from stable MCI in the progression of functional connectivity values over time. However, classifying pMCI and sMCI patients is challenging due to the subtle anatomical differences in the brain~\cite{acharya2019automated}. The four conventional feature extraction approaches usually mentioned in the literature for classifying pMCI and sMCI are voxel-based, slice-based, ROI-based, and patch-based~\cite{zhao2021application}, although they are not entirely mutually exclusive. In this section, before surveying recent advances in deep learning-based methods for classifying pMCI and sMCI, we will briefly review these four approaches by discussing the advantages and disadvantages of each group.

The voxel-based techniques~\cite{basheera2020novel,ruiz20203d,li2020detecting} use the voxel intensity values from all neuroimaging modalities. Although voxel-based techniques are simple to implement, they typically require spatial co-alignment of the input image to standard 3D space and suffer from high dimension feature space compared to available sample numbers. Ortiz et al.~\cite{ortiz2016ensembles} used the \textit{t}-test algorithm to partition the brain area into 3D patches in order to address the mentioned drawbacks and eliminate non-significant voxels. The patches were then used to train an ensemble of deep belief networks, and a voting scheme was used to make the final prediction. However, as mentioned by~\cite{ebrahimighahnavieh2020deep}, there is an inherent over-fitting challenge with voxel-based techniques.

The sliced-based techniques~\cite{zhang2022diagnosis,ebrahimi2021deep} extract slices from the 3D neuroimaging brain scan by projecting the sagittal, coronal, and axial to the 2D image slices. Indeed, because non-affected regions and normal slices must be chosen as the reference distribution, they cannot account for the disease and may be considered an anomaly~\cite{dey2021asc}. Furthermore, choosing separate 2D slices may neglect the spatial dependencies of voxels in adjacent slices due to inter/intra anatomical variances in the brain images~\cite{zhao2021application}. However, sliced-based techniques allow for the usage of a broader range of conventional and deep learning-based approaches. For instance, different pre-trained deep learning models on ImageNet, such as DenseNet, VGG16, GoogLeNet, and ResNet, can be fine-tuned by 2D slices to classify AD from CN~\cite{ebrahimighahnavieh2020deep}. In~\cite{ebrahimi2021deep}, researchers extracted features from MRI image slices using a pre-trained 2D CNN and fed the extracted feature sequence to a recurrent neural network (RNN). The RNN was in charge of determining the relationship between the sequence of extracted features corresponding to MRI image slices. However, sliced-based techniques are computationally expensive due to the use of additional learnable parameters which cannot directly benefit from transfer learning.

ROI-based techniques consider brain regions that have been predefined physically or functionally~\cite{rallabandi2020automatic,liu2020multi,zhao2021region}. These methods use spatial information such as automated anatomical labelling~\cite{tzourio2002automated} and diffusion-weighted imaging in MRI to extract features. The prominent regions that have been considered by almost all ROI-based feature extraction studies on AD prediction are the hippocampus, amygdala, and entorhinal. However, one of the advantages of employing ROI-based approaches is like a double-edged sword which can be a disadvantage because ROI identification requires expert human expertise. Furthermore, these techniques are considered time-consuming due to the need for non-linear registration and brain tissue segmentation. There is also the possibility of information loss because the abnormal region may spread from a single ROI to multiple ROIs.

\begin{figure*}[!b]
    \centering
    \subfloat[]{\includegraphics[width=1\linewidth]{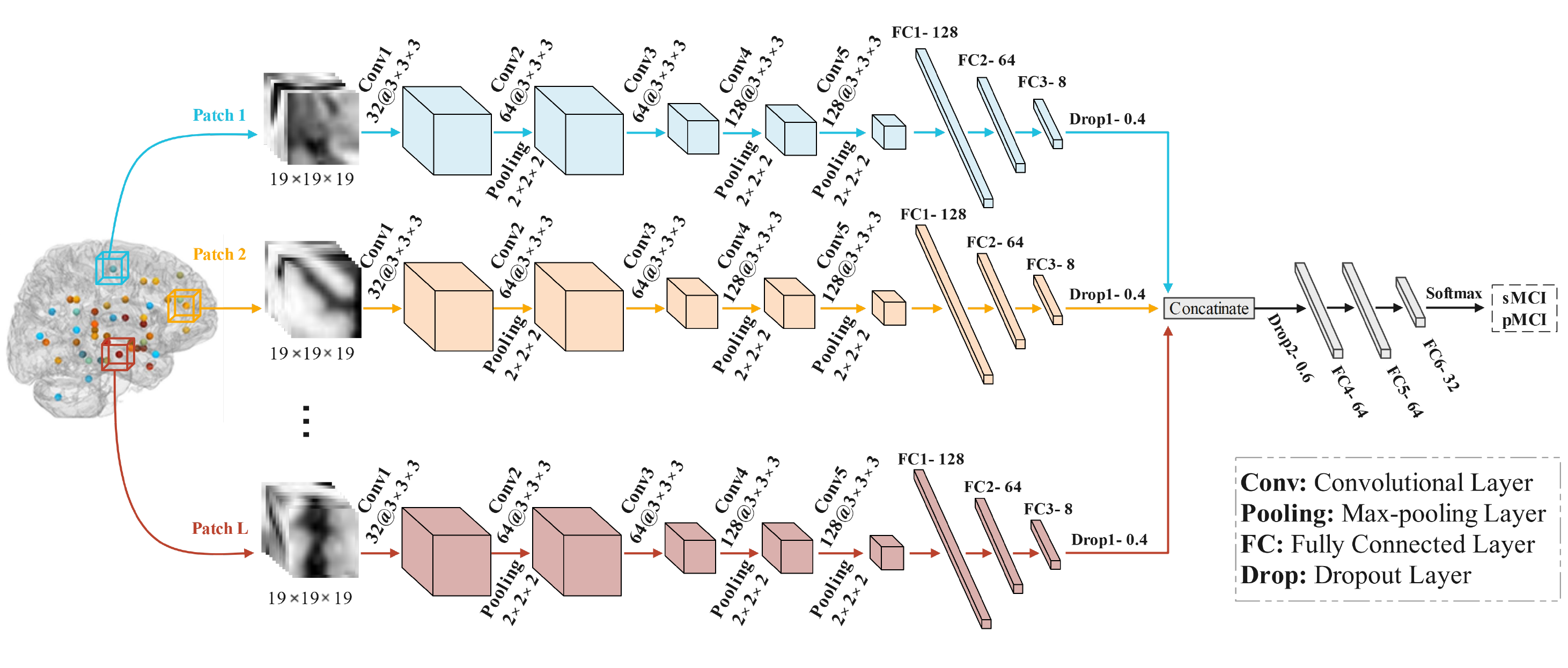}\label{fig:CNN}}
    \caption{The schematic of the proposed multi-stream convolutional neural network.}
    \label{fig:CNN_architecture}
\end{figure*}

Patched-based approaches~\cite{liu2018anatomical,wen2020convolutional} partition the entire brain into multiple patches from which numerous feature vectors are extracted. The extracted patches include shapes, texture, and volume features generated from distinct brain regions or specific patterns. This computational approach eliminates the need for manual ROI identification, and makes it possible to use landmark-based patch extraction or other discriminative patch-based biomarkers~\cite{liu2020multi}. However, selecting useful patches from a complete image is problematic, mainly due to increasing computational complexity when a non-rigid registration approach is used. Researchers have used rigid registration or embedded the registration step into deep learning-based methods to address this challenge~\cite{lian2020attention}. There is another issue with patched-based approaches related to multiple instance learning. Although this challenge is almost addressed in classifying AD and CN by leveraging patch relationships~\cite{liu2018landmark,ilse2018attention}, the difficulty in classifying pMCI in Alzheimer’s disease is not still resolved. This issue is linked to the bag’s labelling in classifying pMCI from sMCI, where a bag can get a negative label even though it contains informative anatomical landmark(s), but it cannot meet the majority rule.

Convolutional neural networks (CNNs) are a popular deep learning approach for classifying Alzheimer's disease. Islam and Zhang~\cite{islam2018brain} proposed three 2D CNNs to generate three distinct MRI views. Each CNN in their architecture comprised three convolutional layers and four dense blocks, where the final decision was made by majority vote. Oh et al.~\cite{oh2019classification} proposed a convolutional autoencoder-based approach for the AD and CN classification task, addressing the pMCI data limitation with transfer learning. Liu et al.~\cite{liu2020multi} suggested a coarse-to-fine hierarchical ensemble learning method for simultaneous hippocampus segmentation and Alzheimer's disease classification that employed a multi-task deep CNN architecture and 3D densely connected CNNs. In this method, an MRI image is first divided into multiple slices, and then a pre-trained deep neural network is used to extract features from the slices. The coarse predictions were then used in ensemble learning to obtain refined results for all slices. Ebrahimi et al.~\cite{ebrahimi2021deep} extracted a sequence of features from 2D MRI slices using a pre-trained ResNet18~\cite{he2016deep}, which they subsequently trained a temporal convolutional network and several types of RNNs to classify AD/CN. Zhao et al.~\cite{zhao2021region} introduced a region ensemble model with three sequential sub-networks to account for a global feature map derived from the entire brain and regional feature maps extracted using a segmentation model. The feature representations were fused in their method, and the classification was performed using an attention-based technique. Researchers employed a data-driven technique to select informative patches in~\cite{liu2020anatomical}, which resulted in specific landmark localisation in brain MRI images. Each landmark patch was then fed into the CNN models, which produced the final classification result using the maximum voting strategy. P{\"o}lsterl et al.~\cite{polsterl2021combining} proposed the dynamic affine feature map transform, an auxiliary module for CNNs that dynamically incites or represses each feature map of a convolutional layer based on both image and tabular biomarkers. A more detailed overview of deep learning algorithms for Alzheimer's disease classification can be found in~\cite{goenka2021deep,zhao2021application,chamarajan2021alzheimer}.

\section{Proposed method}\label{sec:method}

Given a dataset of $N$ samples $\mathcal{D} = \{(x_{i}, y_{i})\}_{i=1}^{N}$, with $x_{i} \in \mathbb{R}^{d_{x}}$ and $y_{i} \in \mathbb{R}^{d_{y}}$, our goal is to train a multi-stream deep convolutional neural network $\mathcal{H}(x) = \mathbb{E}[\mathbf{Y}|\mathbf{X} = x]$ to classify sMCI and pMCI by minimising the cross-entropy between the class labels and the softmax output as in Eq.~\ref{eq:01}

\begin{equation}
    \label{eq:01}
    p(y_i|x; w,b) = \frac{\exp(x^T w_i+b_i) }{\sum_{j\in d_{y}}\exp(x^T w_j+b_j)}.
\end{equation}
where $w$ and $b$ are the network's weights and bias terms, respectively. In this study, we use the baseline 1.5T T1-weighted MRI images of subjects from the ADNI-1 dataset~\cite{jack2008alzheimer}, where the input image $\mathbf{X}$ has a size of $d_{x} = 185 \times 155 \times 150$ and is labelled by $d_{y} = \{0, 1\}$. The output label $\mathbf{Y}$ consists of two probability values in the $[0,1]$ range with $\mathcal{H}(x_{i}) = 0$ if the $i$-th sample belongs to the sMCI class and $\mathcal{H}(x_{i}) = 1$ otherwise.

 It has been shown~\cite{braak1996pattern} that in the early stages of Alzheimer's disease, only certain brain regions are subject to morphological changes caused by the disease. Therefore, we conduct a statistical test to identify these informative landmarks in the MRI images and extract $L$ patches from each MRI image $x_{i} = \{s_{i,j}\}_{j=1}^{L}$, with $s_{i,j} \in \mathbb{R}^{19\times19\times19}$. The proposed data-driven approach for extracting patches from the MRI image, on which a preprocessing step has been performed, is described in the following subsections, followed by the details of the multi-stream CNN. Figure~\ref{fig:CNN_architecture} shows the schematic of the proposed multi-stream architecture.

\subsection{Preprocessing}
\begin{figure}[!htb]
    \centering
    \subfloat[]{\includegraphics[width=0.23\linewidth]{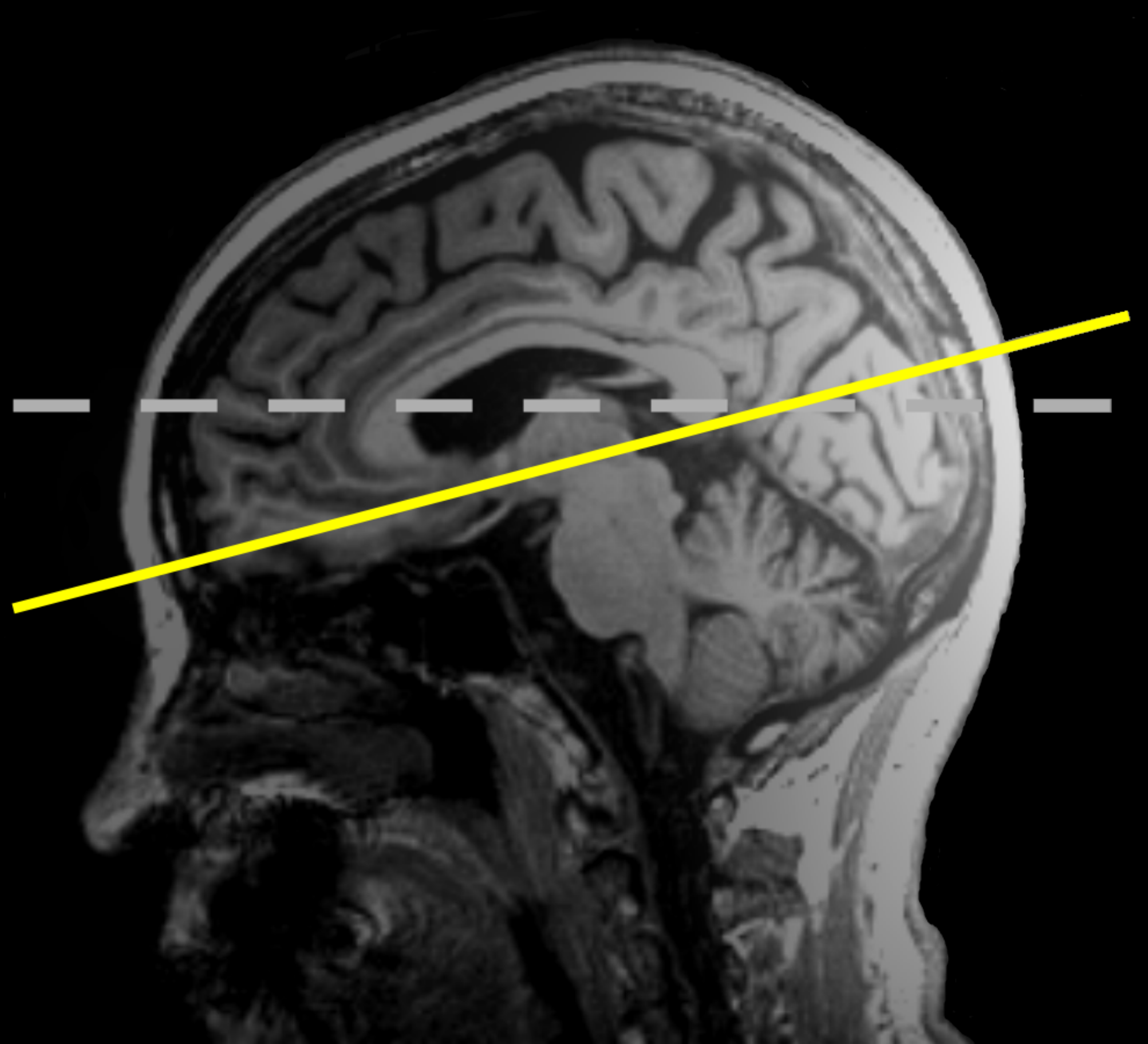}\label{fig:pre1}}\hspace{0.05em}
    \subfloat[]{\includegraphics[width=0.23\linewidth]{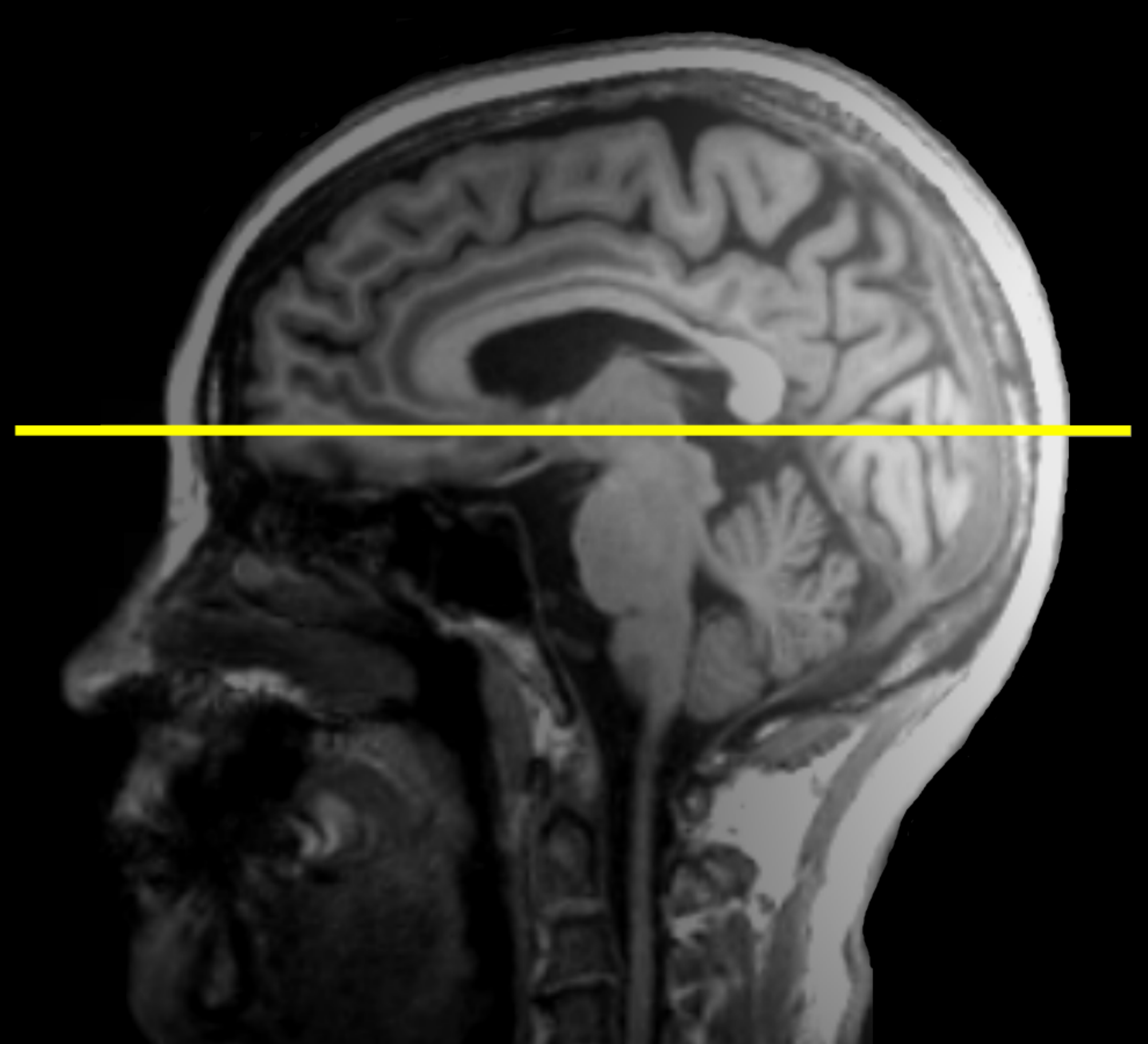}\label{fig:pre2}}\hspace{0.05em}
    \subfloat[]{\includegraphics[width=0.23\linewidth]{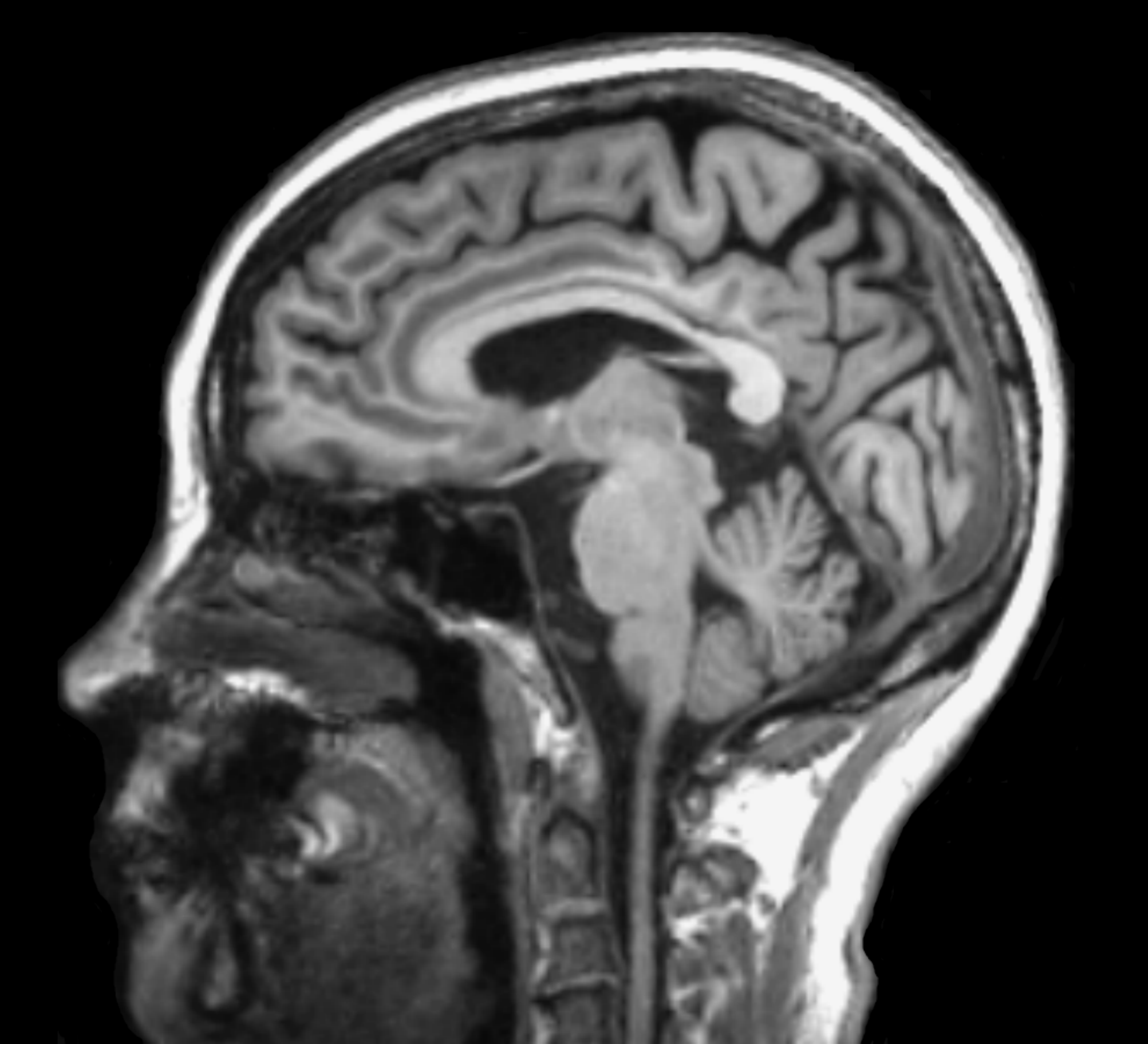}\label{fig:pre3}}\hspace{0.05em}
    \subfloat[]{\includegraphics[width=0.23\linewidth]{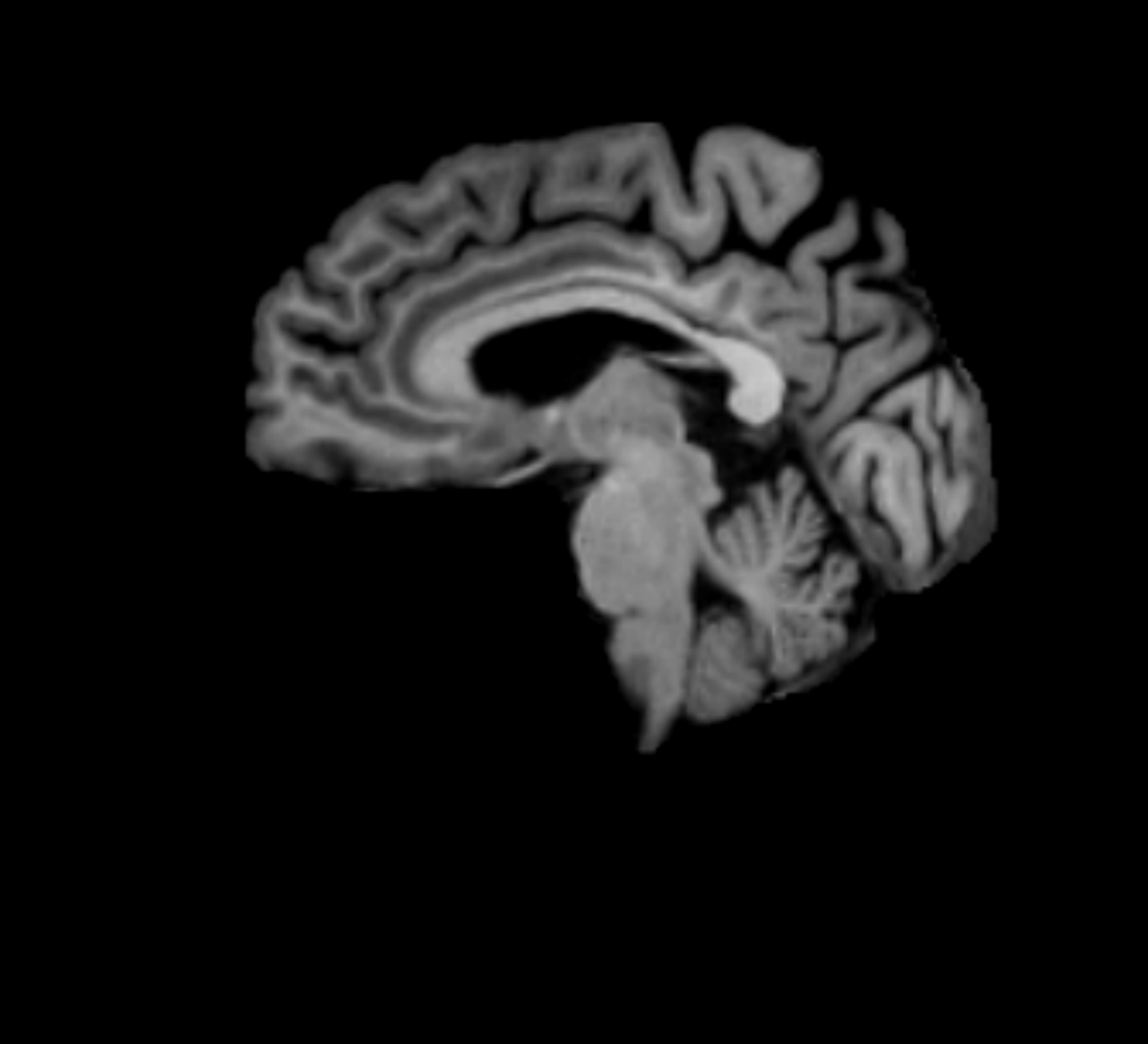}\label{fig:pre4}}\hspace{0.05em}
    \caption{The visual representation of the preprocessing steps for an MRI sample. (a) A raw MRI image, (b) the MRI image with anterior commissure-posterior commissure correction (c) the MRI image with intensity inhomogeneity correction, and (4) skull stripped MRI image. The yellow line in (a) and (b) depicts the anterior commissure-posterior commissure line.}
    \label{fig:preprocess}
\end{figure}

We pre-process the MRI images to use them in the proposed method. There are four steps in the preprocessing phase: (1) anterior commissure-posterior commissure correction using the 3D Slicer software\footnote{http://www.slicer.org/}; (2) intensity inhomogeneity correction using N4ITK~\cite{tustison2010n4itk}, an enhanced version of nonparametric nonuniform normalisation; (3) skull stripping using a pre-trained U-Net\footnote{https://github.com/iitzco/deepbrain} to remove both the skull and the dura; and (4) rigid registration, which involves linearly aligning MRI images to the Colin27 template and resampling them to a size of $155 \times 185 \times 150$ with a resolution of $1 \times 1 \times 1~\mathrm{mm}^{3}$. Figure~\ref{fig:preprocess} shows a sample of MRI image from ADNI-1 dataset on which the preprocessing is performed.

\subsection{Anatomical landmark detection}
We must first identify the anatomical locations in the brain that are most influenced by the disease before we can classify sMCI and pMCI patients. As a result, we randomly divide samples from AD and CN individuals into the train, validation, and test sets with sizes of $0.7 \times N$, $0.1 \times N$, and $0.2 \times N$, respectively, where $N$ is the total number of samples. Then, we select $M = 0.7 \times N$ (\textit{i.e.}, the training set) and propose a novel data-driven landmark detection method in which MRI images are partitioned into $5\times5\times5$ patches. As mentioned in~\cite{fan2008spatial,schroeter2009neural}, when a patient is diagnosed with Alzheimer's disease, several regions in the brain are subject to anatomical degeneration. At the pMCI stage, the same regions undergo anatomical changes, but the degeneration is not as severe as those seen at the onset of Alzheimer's disease. With respect to this fact, we use identical anatomical locations for classifying sMCI and pMCI patients.

\begin{figure}[!htb]
    \centering
    \includegraphics[width=0.8\linewidth]{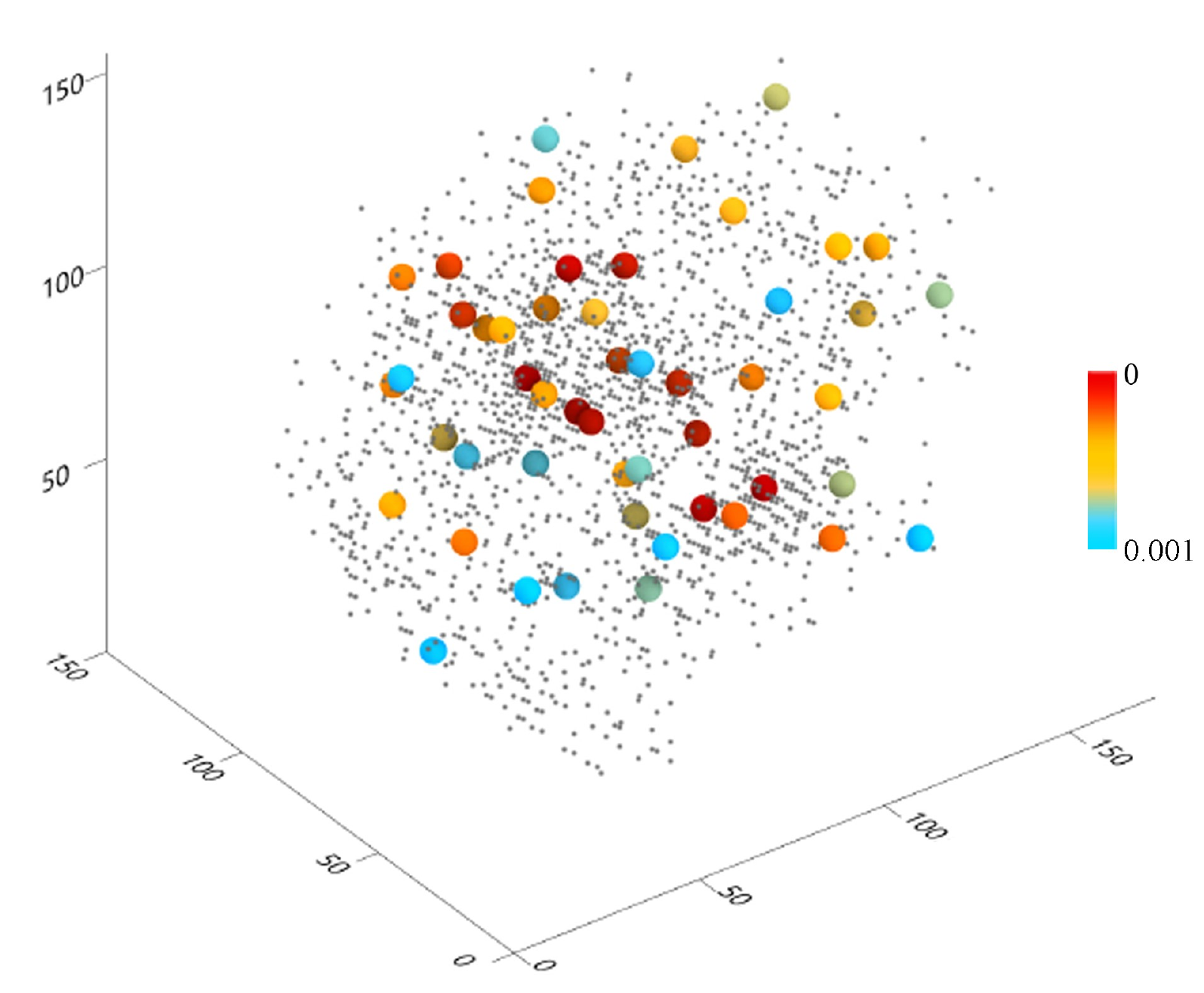}
    \caption{A brain-shaped \textit{p-value} map in which top 50 landmark locations are represented by spheres in a gradient colour from red to blue, with \textit{p-values} ranging from 0 to 0.001. Each \textit{p-value} is paired with a landmark $(l_{x},l_{y},l_{z})$.}
    \label{fig:Landmarks}
\end{figure}

Each partition is then represented by a 29-dimensional feature vector. This feature vector includes the Gray-Level Co-Occurrence Matrix (GLCM)~\cite{haralick1973textural}, Structural Similarity Index Measure (SSIM)~\cite{wang2004image}, Mean Square Error (MSE), entropy, and the mean and standard deviation of the partition voxels. To extract the GLCM elements of the feature vector, we generate six GLCM matrices with three adjacency directions, namely horizontal, vertical, and in-depth, each of which is associated with two distance values. Then, we extract contrast, correlation, homogeneity, and entropy, from each GLCM matrix resulting in a total of 24 elements. The Colin27 template~\cite{holmes1998enhancement} is used as the reference image for measuring SSIM and MSE at each patch location. Finally, we apply the multivariate T2 Hotelling statistical test~\cite{hotelling1992generalization} to generate a brain-shaped \textit{p-value} map (see Fig.~\ref{fig:Landmarks}). Algorithm~\ref{alg:1} details the steps of generating the brain-shaped \textit{p-value} map.

\RestyleAlgo{ruled}
\SetAlgoNoLine
\LinesNumbered
\begin{algorithm}[!htb]
    \SetKwFunction{HT}{Hotelling-Test}\SetKwFunction{FE}{Feature-Extraction}\SetKwFunction{PC}{Partition}\SetKwFunction{ST}{Sort}\SetKwFunction{ED}{Euclidean}
    \SetKwInOut{Input}{Input}\SetKwInOut{Output}{Output}
    \Input{$AD=\{x_{1}^{AD},x_{2}^{AD},\cdots,x_{M}^{AD}\}$\newline
    $CN=\{x_{1}^{CN},x_{2}^{CN},\cdots,x_{M}^{CN}\}$\newline
    $K=34,410$, which is the total number of patches with a size of $5\times5\times5$.}
    \Output{$\mathcal{P}$: A set of \textit{p-value}, forming the brain-shaped \textit{p-value} map.}
    \BlankLine
        \pushline\nonl\textbf{Step 1: Partitioning}\;
        \BlankLine
        \popline $\mathcal{V}^{AD} \gets$ \PC{$AD$}\;
        \tcp{$\mathcal{V}^{AD} = \{p_{1,1},\cdots,p_{1,K}, \cdots,p_{M,1},\cdots,p_{M,K} \}$.}
        $\mathcal{V}^{CN} \gets$ \PC{$CN$}\;
        \tcp{$\mathcal{V}^{CN} = \{q_{1,1},\cdots,q_{1,K}, \cdots,q_{M,1},\cdots,q_{M,K} \}$.}
        \BlankLine
        \pushline\nonl\textbf{Step 2: Feature extraction \& T2 Hotelling test}\;
        \BlankLine
        \popline \For{$j\gets1$ \KwTo $K$}{
            \For{$i\gets1$ \KwTo $M$}{
                $f_{i,j}^{AD} \gets$ \FE{$p_{i,j}$}\;
                $f_{i,j}^{CN} \gets$ \FE{$q_{i,j}$}\;
            }
            $p-value_{j} \gets$ \HT{$f_{M,j}^{AD},f_{M,j}^{CN}$}\;
            \tcp{$f_{M,j}$ is a $M \times 29$ matrix, representing extracted features form the $j^{\text{th}}$ patch.}
        }
        $\mathcal{P} \gets$ \ST{$p-value, \mathrm{asc}$}\;
        \tcp{Each $p-value$ is paired with a landmark $(l_{x},l_{y},l_{z})$.}
    \BlankLine
    \Return{$\mathcal{P}$}
    \caption{Generating brain-shaped \textit{p-value} map.}
    \label{alg:1}
\end{algorithm}

After obtaining $\mathcal{P}$ set, we exclude landmarks with a spatial Euclidean distance of less than 15 to reduce the redundancy of overlapped adjacent patches to identify the most discriminative anatomical landmarks in the brain. The top 50 landmarks with the lowest \textit{p-values} are then chosen (see Fig.~\ref{fig:Landmarks}).

The registration and landmark detection steps in Algorithm~\ref{alg:1} are affected by the MRI image partitioning size, \textit{i.e.}, $5\times5\times5$. In the case of selecting a smaller partition size, the lack of adequate morphological variations could lead to discarding informative landmarks. In contrast, larger partition sizes cause intrinsic physiological differences to eclipse subtle disease-related changes.

Therefore, after obtaining the top 50 landmarks, we sample 27 3D image patches with a $3\times3\times3$ displacement around the centre of each landmark to increase the size of patches to $19\times19\times19$ with two intentions: (i) compensating for regions that may unintentionally be discarded in anatomical landmark detection, and (ii) providing sufficient morphological structures for each stream of the proposed CNN to construct a discriminative latent space.

\subsection{Multi-stream classifier architecture}
We propose a multi-stream CNN architecture with $L$ streams, each fed with the patch $s_{i,j}$ extracted from the input image $x_{i}$ and centred on the identified landmark location $(l_{x,j},l_{y,j},l_{z,j})$. We construct the patch $s_{i,j}$ with a size of $19\times19\times19$, surrounding the corresponding landmark location, in order to better represent morphological variations in the MRI images. The local spectral-spatial feature is extracted from each 3D image patch by each stream of the proposed CNN architecture.

As depicted in Fig.~\ref{fig:CNN_architecture}, the proposed multi-stream CNN has $L=50$ streams, with an identical structure. Each stream has five convolutional layers (Conv), followed by a rectified linear unit (ReLU) activation function. The convolutional layers have 32, 64, 64, 128, and 128 $3\times3\times3$ convolution filters, respectively. After Conv2, Conv4, and Conv5, we consider batch normalisation and $2\times2\times2$ max-pooling layers. There are three fully connected layers with 128, 64, and 8 units at the end of each stream, which are followed by a dropout layer with a ratio of 0.4 to prevent overfitting. Although the architecture of all 50 streams is the same, their weights are tuned and updated separately, where the input patches for each stream are randomly selected to avoid the unintentional bias towards ordering streams. We concatenate the outputs of 50 streams and add a dropout layer with a ratio of 0.6 to fuse the locally extracted spectral-spatial features. Before passing the feature vector into the softmax function for the final classification, we add three fully connected layers with 64, 64, and 32 units, respectively.

\section{Experiments}\label{sec:experiment}
\subsection{ADNI-1 dataset}
\begin{figure}[!htb]
    \centering
    \subfloat[]{\includegraphics[width=0.247\linewidth]{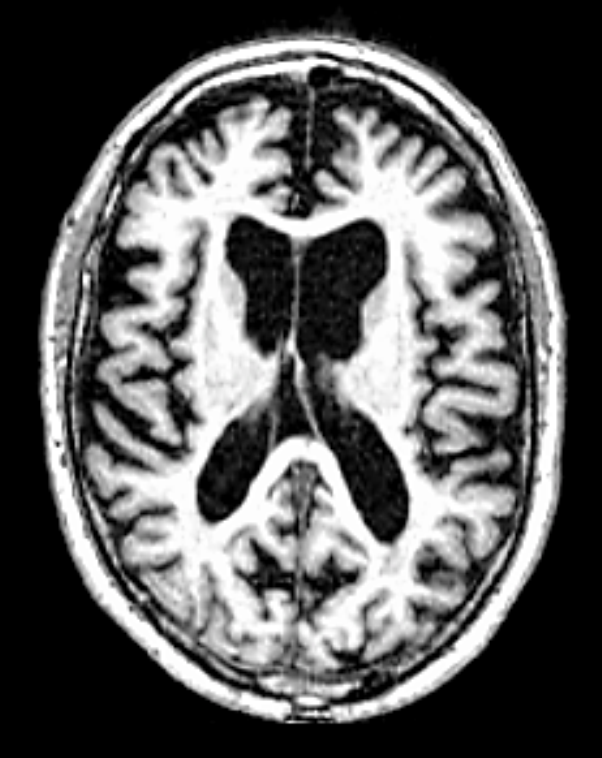}}
    \subfloat[]{\includegraphics[width=0.24\linewidth]{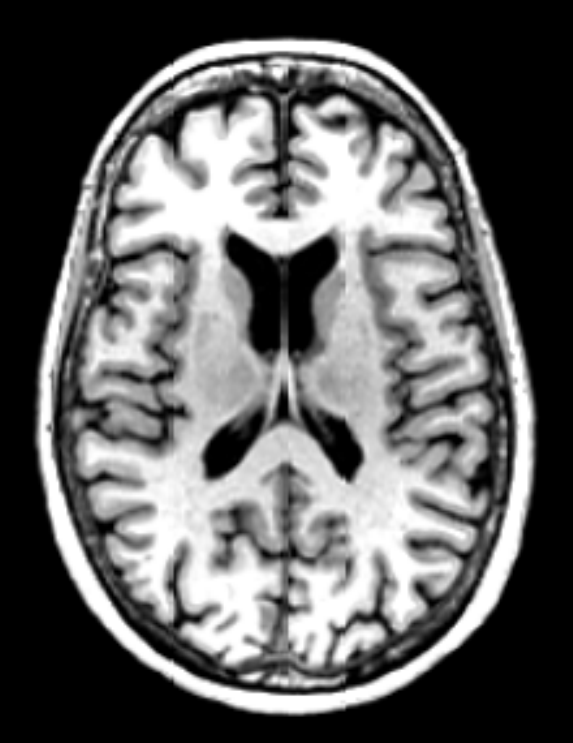}}
    \subfloat[]{\includegraphics[width=0.24\linewidth]{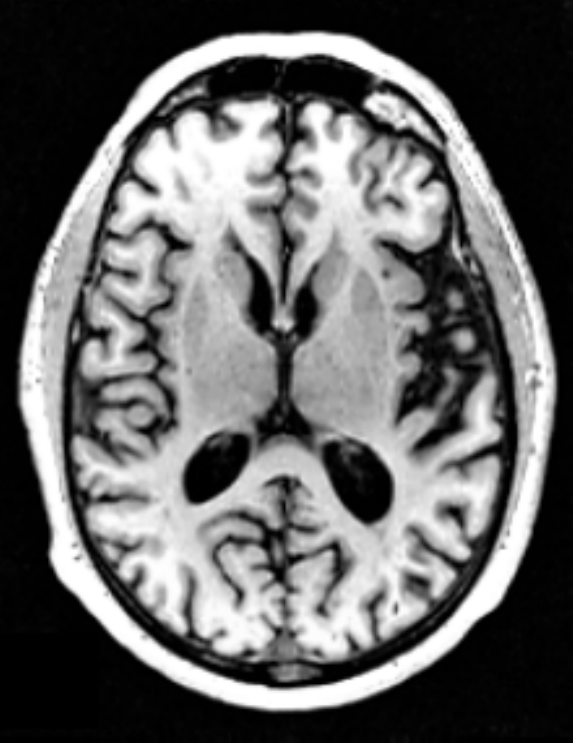}}
    \subfloat[]{\includegraphics[width=0.24\linewidth]{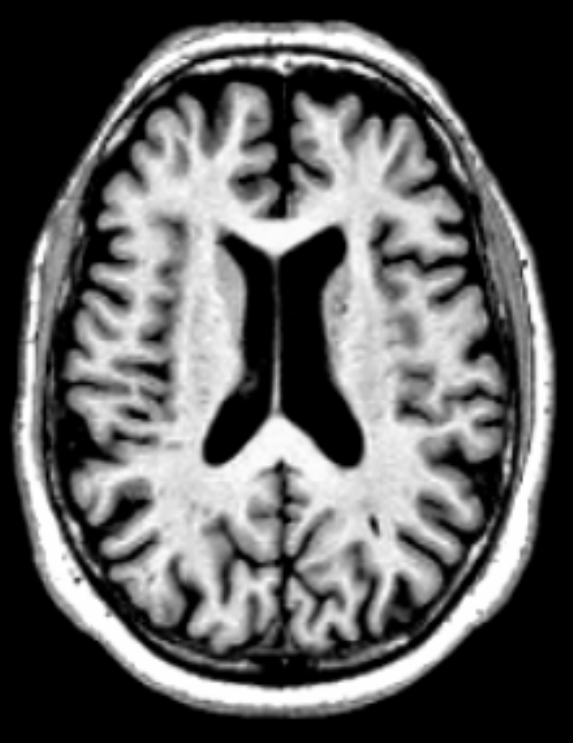}}
    \caption{Four samples from ADNI-1 dataset~\cite{jack2008alzheimer} (a) AD, (b) CN, (c) pMCI, and (d) sMCI}
    \label{fig:samples}
\end{figure}

In this study, we use the baseline 1.5T T1-weighted MRI images of subjects from the ADNI-1 dataset~\cite{jack2008alzheimer}. The volumetric 3D MPRAGE protocol is used to acquire sagittal T1-weighted MRI images with an in-plane spatial resolution of $1.25 \times 1.25~\mathrm{mm}^{2}$ and 1.2~mm thick sagittal slices. The imaging dataset contains baseline images from 695 participants including 200 Alzheimer's disease, 231 cognitively normal, 164 progressive MCI, and 100 stable MCI. Figure~\ref{fig:samples} shows four samples from this dataset, and Table~\ref{table:subjects} presents the demographic and clinical information of subjects in ADNI-1.

\begin{table*}[!htb]
\centering
\caption{Demographic and clinical information of subjects in ADNI-1. Values are reported as mean $\pm$ standard deviation.}
\label{table:subjects}
\resizebox{\textwidth}{!}{%
\begin{tabular}{lccccccc}
\hline
Class & \textbf{Male/Female} & \textbf{Age} & \textbf{ADAS} & \textbf{CDR-sb} & \textbf{FAQ} & \textbf{MMSE} & \textbf{NPI-Q} \\ \hline
AD & 103/97 & 75.64 $\pm$ 7.71 & 13.02 $\pm$ 5.23 & 4.39 $\pm$ 1.6 & 13.16 $\pm$ 6.71 & 23.31 $\pm$ 2.03 & 3.44 $\pm$ 3.27 \\
CN & 119/112 & 76.155 $\pm$ 4.99 & 28.51 $\pm$ 4.89 & 0.03 $\pm$ 0.12 & 0.13 $\pm$ 0.59 & 29.09 $\pm$ 0.98 & 0.36 $\pm$ 0.95 \\
sMCI & 66/34 & 75.44 $\pm$ 7.27 & 22.93 $\pm$ 5.78 & 1.24 $\pm$ 0.62 & 1.65 $\pm$ 3.00 & 27.65 $\pm$ 1.70 & 1.45 $\pm$ 2.40 \\
pMCI & 97/67 & 74.54 $\pm$ 7.05 & 17.67 $\pm$ 5.14 & 1.87 $\pm$ 0.96 & 5.64 $\pm$ 5.15 & 26.62 $\pm$ 1.71 & 2.30 $\pm$ 3.11 \\ \hline
\multicolumn{4}{l}{\begin{tabular}[c]{@{}l@{}}\textbf{ADAS:} Alzheimer's Disease Assessment Scale\\ \textbf{CDR-sb:} Clinical Dementia Rating `sum of boxes’\\ \textbf{FAQ:} Functional Activities Questionnaire\end{tabular}} & \multicolumn{4}{l}{\begin{tabular}[c]{@{}l@{}}\textbf{MMSE:} Mini-Mental State Examination\\ \textbf{NPI-Q:} Neuropsychiatric Inventory Questionnaire\end{tabular}}
\end{tabular}%
}
\end{table*}

\subsection{Architecture details and Evaluation metrics} \label{subsec:architecture}
The proposed architecture is implemented using Python based on the Keras package \footnote{https://github.com/fchollet/keras}, on a computer with Intel(R) Core(TM) i7-4790K @4.00 GHz CPU and 16G RAM. We trained the network using Adam optimiser~\cite{kingma2014adam} with the first momentum of 0.9 and the second momentum of 0.999. The initial learning rate and the constant for numerical stability are set to $10^{-3}$ and $10^{-6}$, respectively. We set the maximum number of training epochs to 40 and used a mini-batch-size of 5 at each iteration, where the training data was shuffled before each training epoch. There are two other hyper-parameters which are the number of streams $L$ and the patch size. 

\begin{figure}[!htb]
    \centering
    \subfloat[]{\includegraphics[width=0.47\linewidth]{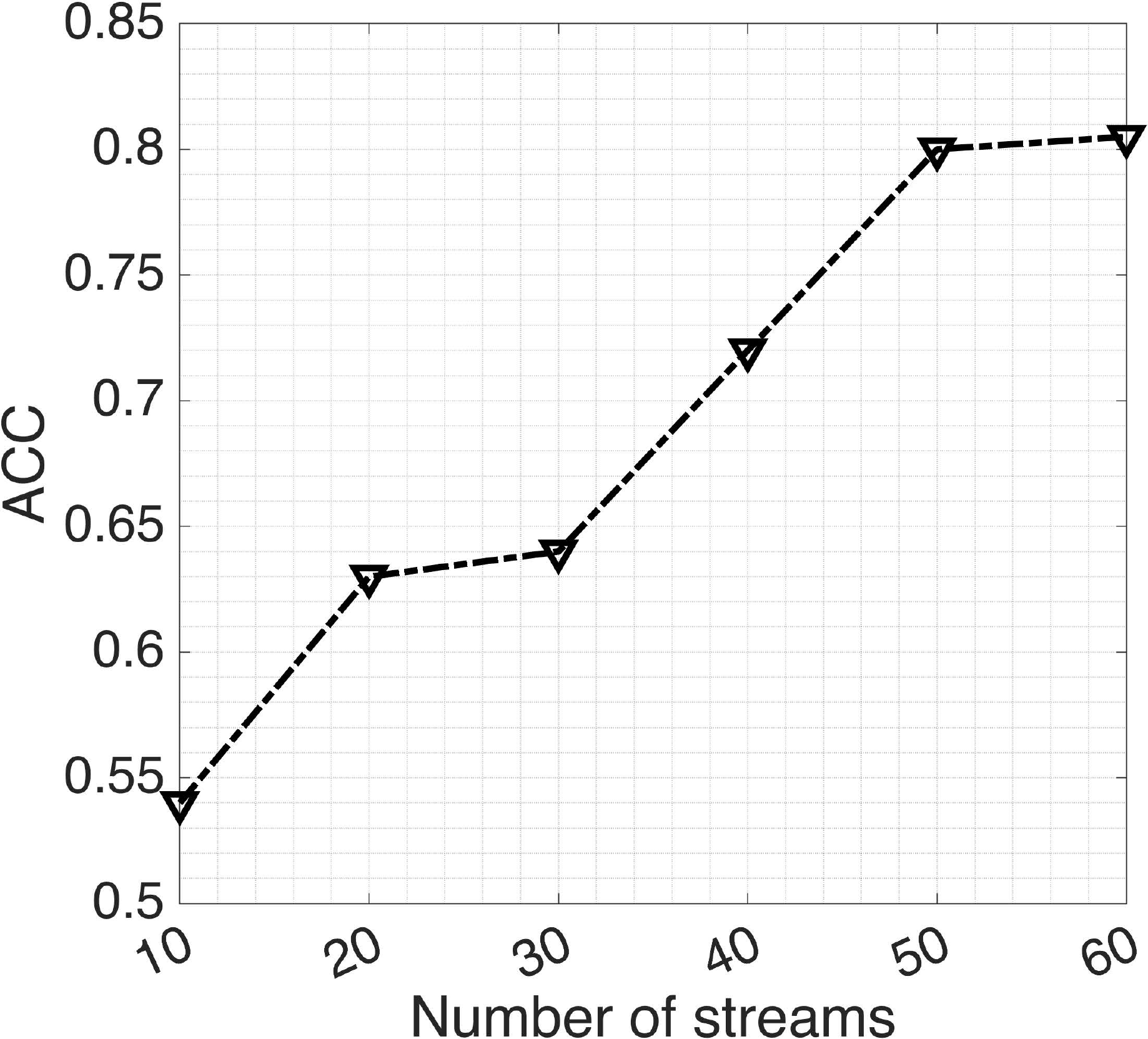}\label{fig:patches_no}}\quad
    \subfloat[]{\includegraphics[width=0.47\linewidth]{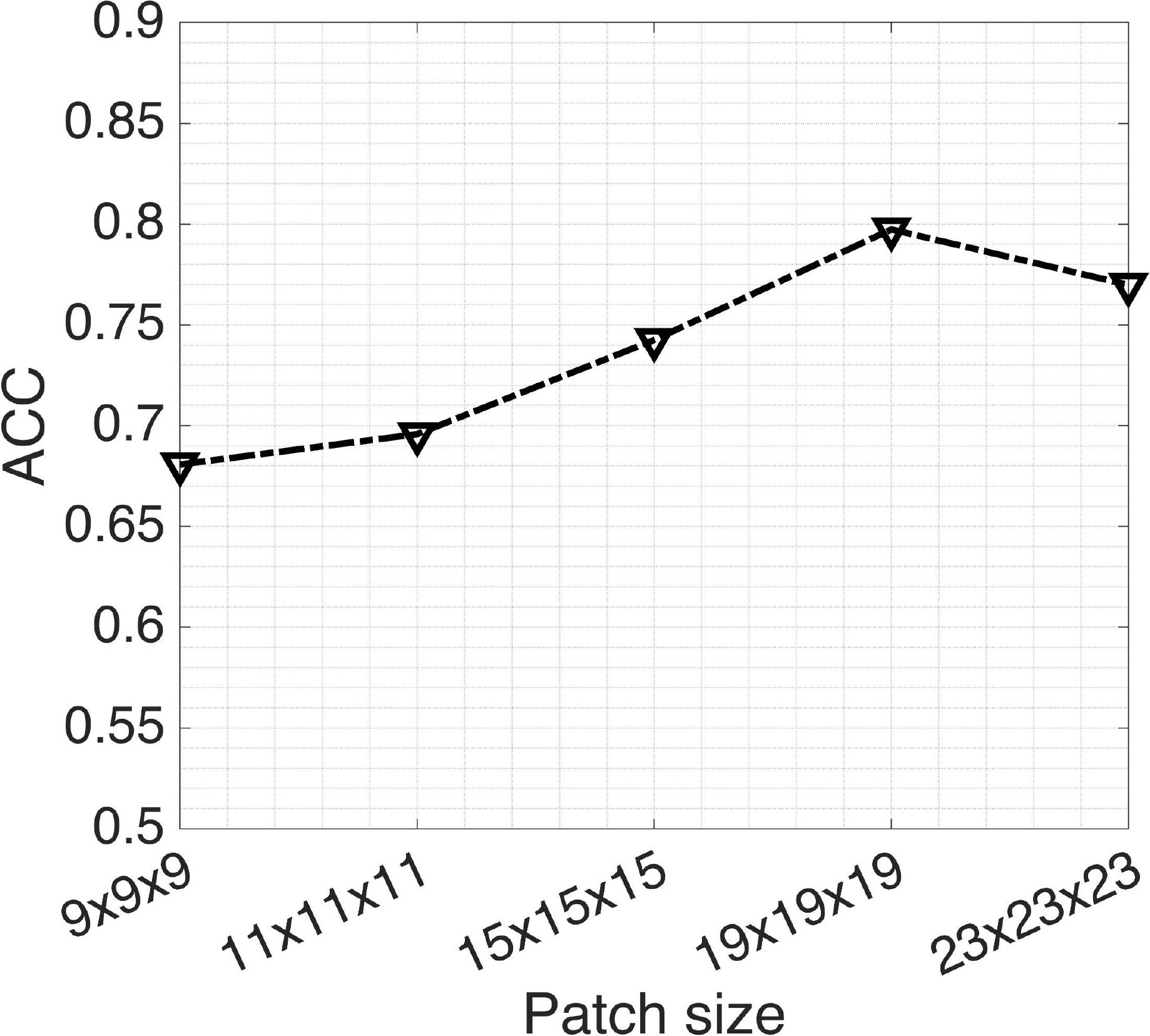}\label{fig:patches_size}}
    \caption{The accuracy of the proposed multi-stream CNN classifier as a function of (a) the number of streams, and (b) the patch size.}
    \label{fig:parameter}
\end{figure}

We evaluate the performance of the proposed architecture with the number of streams in the range $\{10, 20, 30, 40, 50, 60\}$ and the size of patches in the range $\{9, 11, 15, 19, 23\}$ using the validation set from the AD and CN classes.It is worth noting that because there are two hyperparameters, one must be fixed in order to generate the other. Figure~\ref{fig:parameter} shows the accuracy of the proposed multi-stream CNN as a function of these two hyper-parameters.

While increasing the number of streams improves classification accuracy (see Fig.~\ref{fig:patches_no}), it also increases computational complexity. We can also observe in Fig.~\ref{fig:patches_size} that a larger patch size (\textit{i.e.}, $23 \times 23 \times 23$) shows a performance drop when compared with the $19 \times 19 \times 19$ patch size. This is owing to the fact that enlarging the patch size includes tissues in that patch of the brain whose intrinsic morphological similarity is dominant to subtle changes that aid in the early diagnosis of the disease. Therefore, instead of discriminating the subtle changes induced by the disease's development in the early stages, the network discriminates irrelevant tissues.As a result, to establish a trade-off between computational complexity and accuracy, we set the number of streams $L$ to 50 and the patch size to $19 \times 19 \times 19$ and preserve these values in the rest of the experiments.

To evaluate the performance of the proposed architecture, we use the metrics as in Eq.~\ref{eq:02}.
\begin{equation}
    \label{eq:02}
    \begin{split}
        \mathrm{ACC} = &~\frac{TP+TN}{TP+TN+FP+FN}. \\ 
        \mathrm{SEN} = &~\frac{TP}{TP+FN}, \\
        \mathrm{SPE} = &~\frac{TN}{TN+FP}, \\
        \mathrm{F1-score} = &~\frac{2\times \mathrm{SEN}\times \mathrm{SPE}}{\mathrm{SEN}+\mathrm{SPE}}.
    \end{split}
\end{equation}
where $TP$, $TN$, $FP$, and $FN$ denote true positive, true negative, false positive, and false negative, respectively. We also report area under receiver operating characteristic curve (AUC).

\subsection{Experimental results}
To assess the performance of the proposed architecture, we performed three steps: (1) training the proposed multi-stream CNN with MRI images of AD and CN subjects, (2) transferring weights from the trained model in Step 1 to the identical architecture and fine-tuning it with the data related to sMCI and pMCI patients, and (3) adding biomarkers as an auxiliary modality to examine the reciprocal influence of spectral-spatial features and biomarkers.\\

\noindent\textbf{(1) Classification of AD and CN subjects}: To train the proposed multi-stream CNN, we randomly select 70\% of the MRI patches from the two classes of AD and CN as the training set, 20\% as the test set, and 10\% as the validation set. We also compare the trained model in this step with 10 additional approaches, including regions of interest (ROI)~\cite{zhang2011multimodal}, Voxel-Based Morphometry (VBM)~\cite{ashburner2000voxel}, and five deep learning-based methods. We used the FAST algorithm~\cite{smith2004advances} to segment brain MRI images into three different tissues for the ROI and VBM comparison: White Matter (WM), Gray Matter (GM), and Cerebrospinal Fluid (CSF). We followed the implementations described by the researchers for the deep learning-based approaches.

\begin{table*}[!htb]
\centering
\caption{Results of classifying AD/CN and pMCI/sMCI using the proposed multi-stream CNNs along with 10 additional approaches.}
\label{table:results}
\resizebox{\textwidth}{!}{%
\begin{tabular}{lccccccccccc}
\hline
\multirow{2}{*}{Method} & \multicolumn{5}{c}{\textbf{AD vs. CN (\%)}} & \multicolumn{1}{l}{} & \multicolumn{5}{c}{\textbf{pMCI vs. sMCI (\%)}} \\ \cline{2-6} \cline{8-12} 
 & ACC & SEN & SPE & F1-score & AUC &  & ACC & SEN & SPE & F1-score & AUC \\ \hline
ROI + SVM & 70.30 & 68.51 & 67.50 & 67.83 & 79.39 &  & 59.47 & 66.86 & 68.90 & 67.87 & 60.99 \\
VBM + SVM & 82.84 & 82.54 & 80.50 & 81.34 & 90.39 &  & 70.08 & 81.48 & 67.07 & 73.58 & 74.38 \\
ROI + MLP & 73.08 & 70.00 & 71.19 & 70.59 & 77.52 &  & 58.75 & 70.00 & 66.04 & 67.96 & 53.93 \\
VBM + MLP & 83.85 & 78.33 & 85.45 & 81.74 & 89.92 &  & 70.00 & 72.00 & 78.26 & 75.00 & 73.93 \\
Shmulev et al.~\cite{shmulev2018predicting} & - & - & - & - & - &  & 76.00 & 70.00 & 88.00 & 77.97 & 86.00 \\
$\mathrm{DM}^{2}\mathrm{L}$~\cite{liu2019joint} & 91.09 & 93.50 & 88.05 & 90.69 & 95.86 &  & 76.90 & 82.43 & 42.11 & 55.74 & 77.64 \\
H-FCN~\cite{lian2020hierarchical} & 90.30 & \textbf{96.50} & 82.40 & 85.08 & 95.10 &  & 80.90 & \textbf{85.40} & 52.60 & 65.10 & 78.10 \\
HybNet~\cite{lian2020attention} & 91.90 & 82.40 & 94.50 & 91.50 & 96.50 &  & 82.70 & 57.90 & 86.60 & 69.39 & 79.30 \\
Zhao et al.~\cite{zhao2021region} & - & - & - & - & - &  & \textbf{85.90} & 50.00 & 91.60 & 64.68 & 85.40 \\
Proposed architecture $\bigoplus$ biomarkers & 97.54 & 95.54 & 99.40 & 97.43 & 99.38 &  & 69.21 & 68.56 & 93.15 & 78.98 & 77.35 \\
Proposed architecture & \textbf{97.78} & 95.59 & \textbf{99.82} & \textbf{97.66} & \textbf{99.97} &  & 79.90 & 75.55 & \textbf{99.70} & \textbf{85.96} & \textbf{94.39} \\ \hline
\end{tabular}%
}
\end{table*}

For the RIO-based method, we align the anatomical automated labelling template~\cite{tzourio2002automated} with the native space of each MRI image. This template contains 116 predefined regions in which the extracted GM tissue volumes are normalised using the summation of GM, WM, and CSF volumes. This normalised tissue volume is used as the representative feature for the MRI images. For the VBM method, we use affine registration with 12 degrees of freedom to align MRI images with the Colin27 template~\cite{holmes1998enhancement} in order to extract the GM density as the representative feature. To reduce the dimensionality of this feature vector, we perform the \textit{t}-student statistical test on the extracted features from AD and CN subjects and chose GM densities with \textit{p-values} less than 0.001.

Finally, we classify AD and CN using these two feature representations using a linear support vector machine (SVM)~\cite{cortes1995support} with soft-margin and a multilayer perceptron (MLP). The MLP comprises two hidden layers, with 13 and 15 neurons, and a binary output layer with a logistic activation function. We train these classifiers using the 5-fold cross-validation strategy. To ensure a fair comparison for MCI conversion prediction, we train the VBM and ROI models on AD/CN and test them on an independent test set of pMCI/sMCI.\\

\noindent\textbf{(2) Classifying sMCI and pMCI using transfer learning}: The availability of MRI images for sMCI and pMCI patients is less than that of AD because the symptoms of Alzheimer's disease are not as severe at these stages. As a result, many patients may not be asked to get an MRI test. Furthermore, structural changes in MCI brains caused by dementia may be very subtle compared to CN and AD, making the convergence of the proposed multi-stream CNN challenging. To overcome these limitations, we first trained our model using data from CN and AD MRI images and fine-tune the model with data gathered from sMCI and pMCI patients.

Since the pre-trained streams of the proposed architecture are designed to extract structural features that correspond to AD, we freeze the initial layers and only re-train the last three fully connected layers. To fine-tune the model, we used 70\% of the pMCI and sMCI MRI patches and tested the fine-tuned model with the remaining data.\\

\noindent\textbf{(3) Unleashing the impact of biomarkers}: In addition to MRI images, non-invasive biomarkers such as demographic information and cognitive test scores~\cite{hett2021multi} can also be used to provide potentially discriminatory information for diagnosing Alzheimer's disease in its early stages. We propose adding biomarkers as an auxiliary modality to the architecture in order to evaluate the reciprocal influence of spectral-spatial features and biomarkers. Therefore, we introduce an additional input stream to the proposed architecture to incorporate numerical values of \{Age, ADAS, CDR-sb, FAQ, MMSE, NPI-Q\} (see Table~\ref{table:subjects}). However, a subset of the MRI images in the ADNI-1 dataset does not include biomarkers, resulting in missing values when incorporating them into the proposed CNN architecture. We use $k$-Nearest Neighbour (kNN) as a preprocessing step to handle missing values. First, we execute the kNN with $k = 6$ (analogous to 6 biomarkers) on the available data. Then, for those entries that do not have a biomarker, we fill in the value with the average values of the $k$ nearest neighbours. Finally, we incorporate these biomarkers into the architecture at the `Concatenate' layer while preserving the learning parameters as described in Section~\ref{subsec:architecture}.

\begin{figure}[!htb]
    \centering
    \subfloat[]{\includegraphics[width=0.45\linewidth]{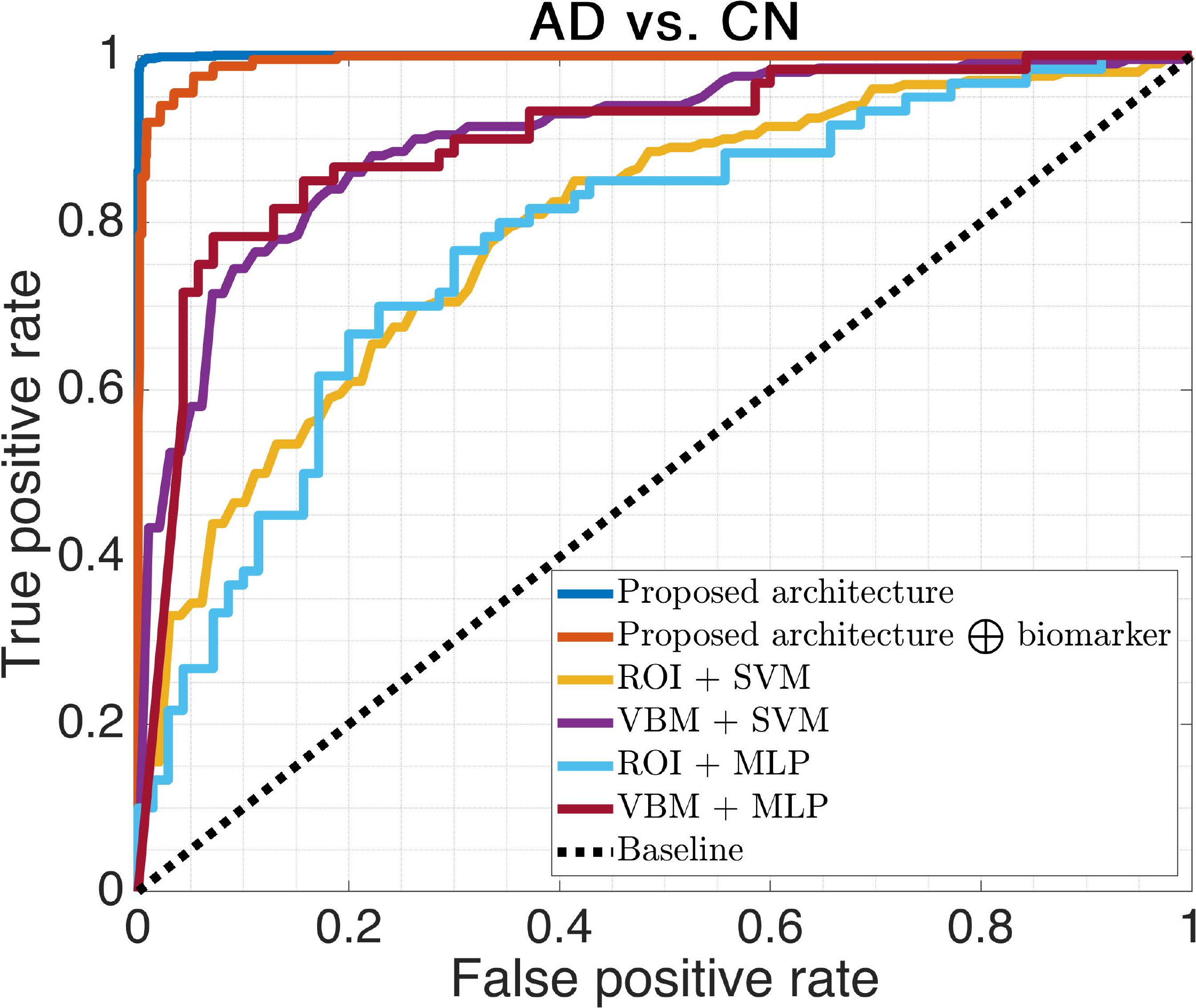}\label{fig:ADCN_ROC}}\quad
    \subfloat[]{\includegraphics[width=0.45\linewidth]{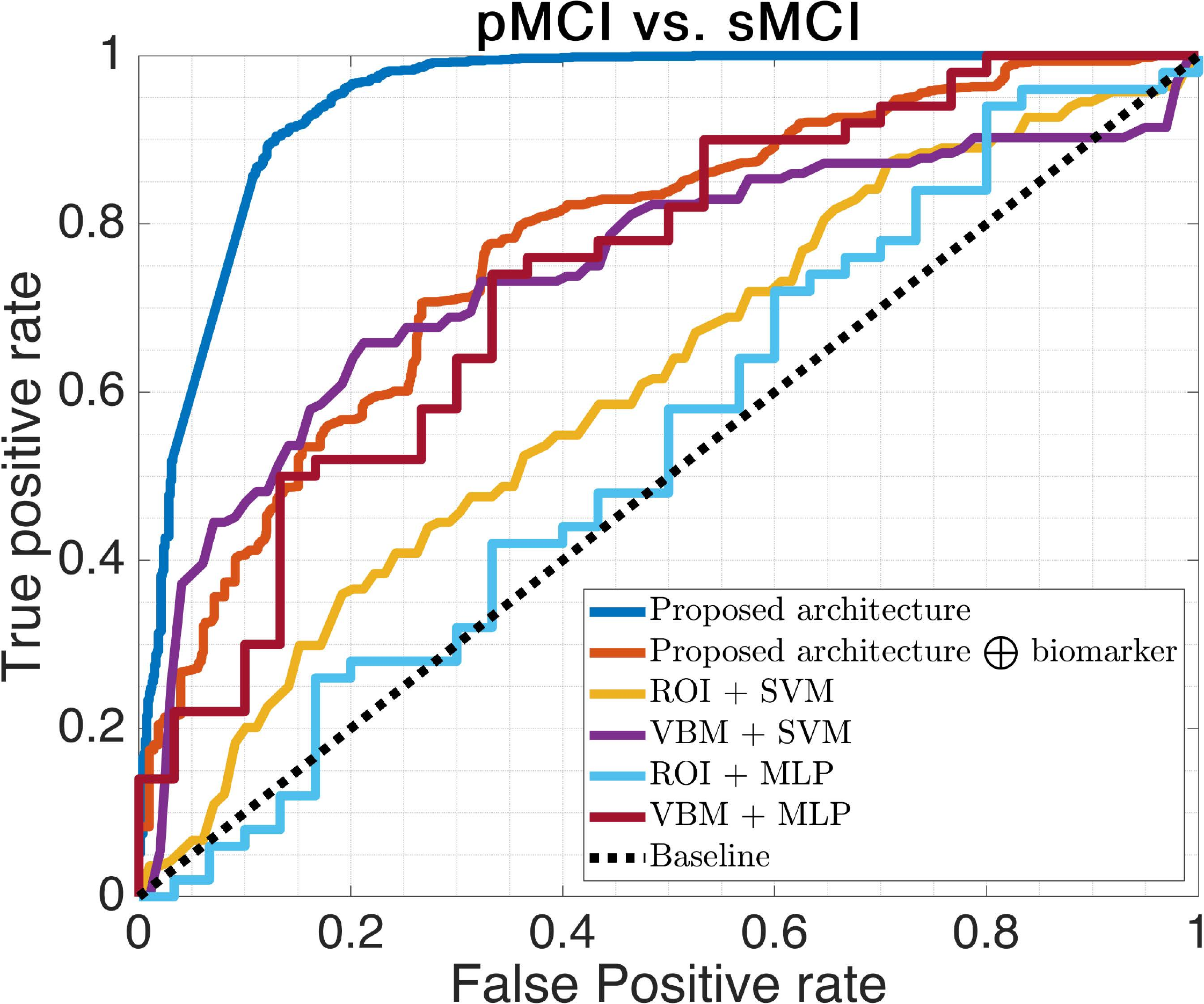}\label{fig:MCI_ROC}}
    \caption{ROC curves of the proposed multi-stream architecture, the proposed architecture integrated with biomarkers, ROI, and VBM-based methods for (a) AD/CN classification, and (b) pMCI/sMCI classification.}
    \label{fig:ROC_curves}
\end{figure}

Table~\ref{table:results} shows the performance of the proposed architectures compared with 10 other approaches. As shown in Table~\ref{table:results}, the proposed multi-stream CNNs achieve considerably better F1-scores than both conventional and deep learning-based approaches. While the biomarkers can improve the classification performance, the lack of large-scale data is a barrier to incorporating them into the proposed method. Furthermore, as previously stated, the cognitive test scores of individuals suffering from various types of MCI differ slightly. This subtle difference could explain why the multi-stream architecture combined with biomarkers performs poorly in distinguishing pMCI and sMCI when compared with the AD and CN classification. We also plot the ROC curves in Fig.~\ref{fig:ROC_curves} for classifying both AD vs. CN and pMCI vs. sMCI.

\begin{figure*}[!htb]
    \centering
    \subfloat[]{\includegraphics[width=0.55\linewidth]{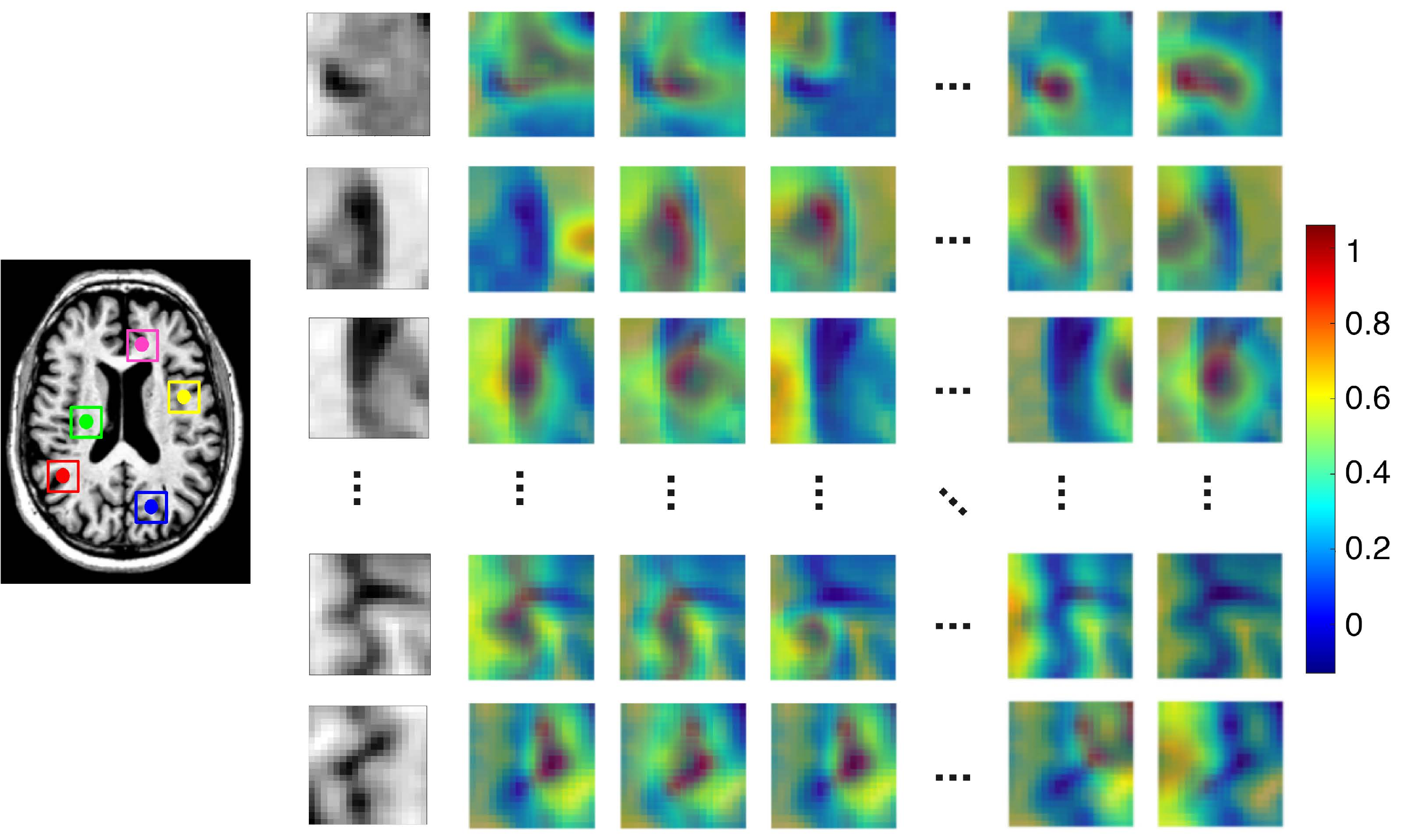}\label{fig:gradcam}}\quad
    \subfloat[]{\includegraphics[width=0.41\linewidth]{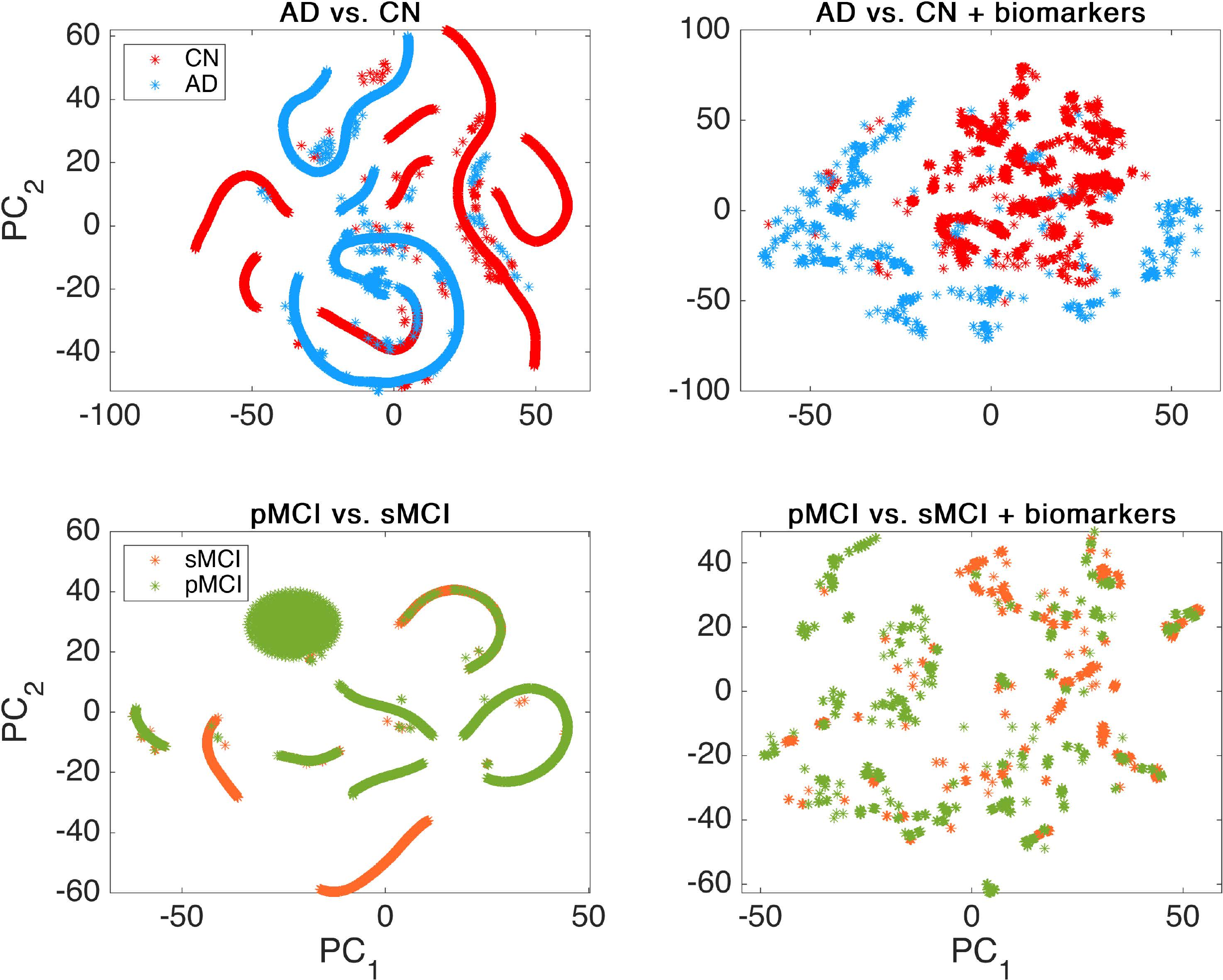}\label{fig:tsne}}
    \caption{(a) An example of a class-discriminative localisation map for the last convolutional layers in the proposed multi-stream CNN using Grad-CAM~\cite{selvaraju2017grad}. Each stream has 128 filters. However, in order to keep the publication page limits, we only illustrate five out of 50 patches, each representing five filters. The colourbar represents the relevance of each pixel in the Grad-CAM, where the red colour has the highest importance in the trained network's final decision and the blue colour has the least importance. (b) Visualisation of the feature space using t-SNE~\cite{hinton2002stochastic} for the classification of AD/CN and pMCI/sMCI with and without biomarkers.}
    \label{fig:feature}
\end{figure*}

In the case of classifying AD and CN, Fig.~\ref{fig:ADCN_ROC} shows that the proposed architecture achieves an AUC of 99.97\%, outperforming ROI, VBM, and multi-stream architecture integrated with biomarkers, which achieve AUCs of 79.39\%, 90.39\%, and 99.38\%, respectively. Furthermore, as demonstrated in Fig.~\ref{fig:MCI_ROC}, when we use transfer learning, the fine-tuned model can provide appropriate discriminative information to classify pMCI and sMCI subjects. This evidence can also convey that AD and pMCI have structural deformations similar to the extracted features from various landmarks in different streams of the proposed architecture.

The results in Table~\ref{table:results} show that the proposed multi-stream architecture combined with biomarkers is slightly less efficient than the multi-stream CNN in distinguishing pMCI from sMCI when compared with the AD and CN classification. To highlight these subtle differences and better comprehend the reciprocal influence of spectral-spatial features and biomarkers, we have  visualised the class-discriminative localisation map and the latent space in Fig.~\ref{fig:feature} using Grad-CAM~\cite{selvaraju2017grad} and t-SNE~\cite{hinton2002stochastic}, respectively. Figure~\ref{fig:gradcam} shows an example of a class-discriminative localisation map for the last convolutional layers in the proposed multi-stream CNN in which the salient regions of each brain patch have been identified so that the classifier could distinguish between AD/CN and pMCI/sMCI with high potential. Furthermore, we have observed in Fig.~\ref{fig:tsne} that introducing biomarkers as an auxiliary modality causes the clusters (\textit{i.e.}, AD/CN and pMCI/sMCI) to become tighter with a degree of overlap than without the biomarkers. This evidence can explain why the proposed multi-stream CNN without biomarkers performs marginally better, where the latent space provides for clean cluster separation.

\subsection{Ablation study}
\label{subsec:ablation}
In the ablation study, we conduct two sets of experiments to better understand the efficacy of the proposed multi-stream CNN for the classification of pMCI in Alzheimer's disease. We therefore examine:

\noindent\textbf{--- Contribution of transfer learning}: In this experiment, instead of using transfer learning, we test the baseline performance when the model is directly trained on pMCI/sMCI data. We have 100 and 164 sMCI and pMCI samples, respectively, and use the same data split as in the transfer learning (70/30) for a fair comparison. The other learning parameters are preserved as described in Section~\ref{subsec:architecture}. The evaluation metrics (\textit{i.e.}, ACC: 63.53\%, SEN: 100\%, SPE: 0.0\%, and F1-score: 77.70\%) reveal that the model cannot converge to classify pMCI/sMCI classes without using transfer learning. In fact, the trained model predicts the same optimal output regardless of the input, with the average answer minimising loss.

\noindent\textbf{--- Contribution of multi-stream architecture}: In this experiment, instead of the proposed multi-stream CNN, we use a single stream CNN to which 3D patches are fed. The model is trained on the AD/CN dataset and fine-tuned using pMCI/sMCI data with the same data split used in the previous experiments. The evaluation metrics (\textit{i.e.}, ACC: 56.96\%, SEN: 67.35\%, SPE: 40.00\%, and F1-score: 66.00\%) show that the model is unable to converge to classify pMCI/sMCI classes, and the trained model is predicting random class for all the data points regardless of the input. These metrics highlight two main facts: (1) with the multi-stream architecture, the model can build a more discriminative latent space as a result of focusing on the landmarks identified in the anatomical landmark detection phase, which are more likely to be signs of Alzheimer's disease. However, with a single-stream architecture, the model cannot distinguish between anatomical abnormalities caused by the disease and those caused by the morphology of the brain. Indeed, the model's sensitivity and specificity are inadvertently changed due to the morphological similarity of the inputs; (2) input patches to the single-stream architecture predominantly capture local information of the image, and the global relationship between different landmark locations is no longer taken into account.

\section{Conclusion}\label{sec:conclusion}
Alzheimer's disease is the most common type of dementia, resulting in memory impairment and cognitive decline. Mild cognitive impairment is a prodromal stage of AD, also known as the transition stage. MCI patients either progress to AD or remain at the same stage over time. Therefore, it is critical to distinguish between progressive MCI and stable MCI in early stages to prevent rapid progression of the disease.

In this study, we have proposed a method for classifying pMCI and sMCI patients using MRI images. The proposed method consists of two main steps. (1) We have developed a novel data-driven approach based on the multivariate T2 Hotelling statistical test to identify anatomical landmarks in MRI images and generate a brain-shaped \textit{p-value} map. Each landmark is paired with a 3D coordinate, allowing us to extract patches of $19\times19\times19$. (2) We have proposed a multi-stream deep convolutional neural network in which each stream is fed by one of the patches. This multi-stream CNN employed transfer learning to classify pMCI and sMCI patients using the ADNI-1 dataset. We assessed the proposed method in three experimental steps and demonstrated the significance of our contributions to transfer learning and multi-stream architecture in the Ablation study. We have performed several experiments to evaluate the efficiency of the proposed architecture based on the best practices. Experimental results have shown that our method outperformed existing approaches, particularly in the classification of MCI patients. Thus, this method can assist practitioners to expand investigating on various diseases associated with structural atrophy.

{
\bibliographystyle{IEEEtran.bst}
\bibliography{references}
}

\vspace*{-2\baselineskip}
\begin{IEEEbiography}[{\includegraphics[width=1in,height=1.25in,clip,keepaspectratio]{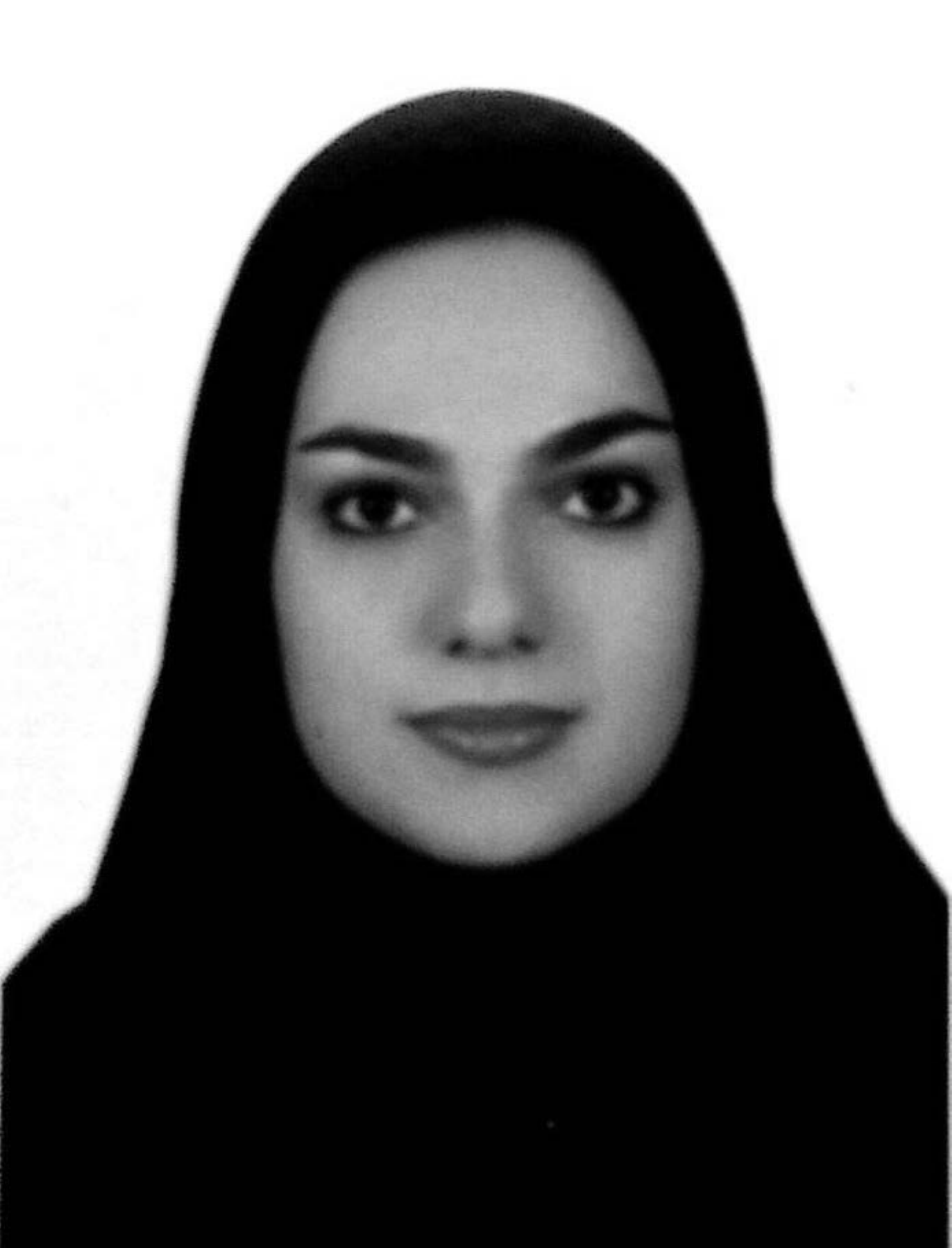}}]{Mona Ashtari-Majlan} received her Master’s degree in Health Systems Engineering from Amirkabir University of Technology, Tehran, in 2021. She is a PhD candidate in computer science at Universitat Oberta de Catalunya, Spain. Her area of interest includes Biomedical Image Processing, Computer Vision, and Deep Learning.
\end{IEEEbiography}

\vspace*{-2\baselineskip}
\begin{IEEEbiography}[{\includegraphics[width=1in,height=1.25in,clip,keepaspectratio]{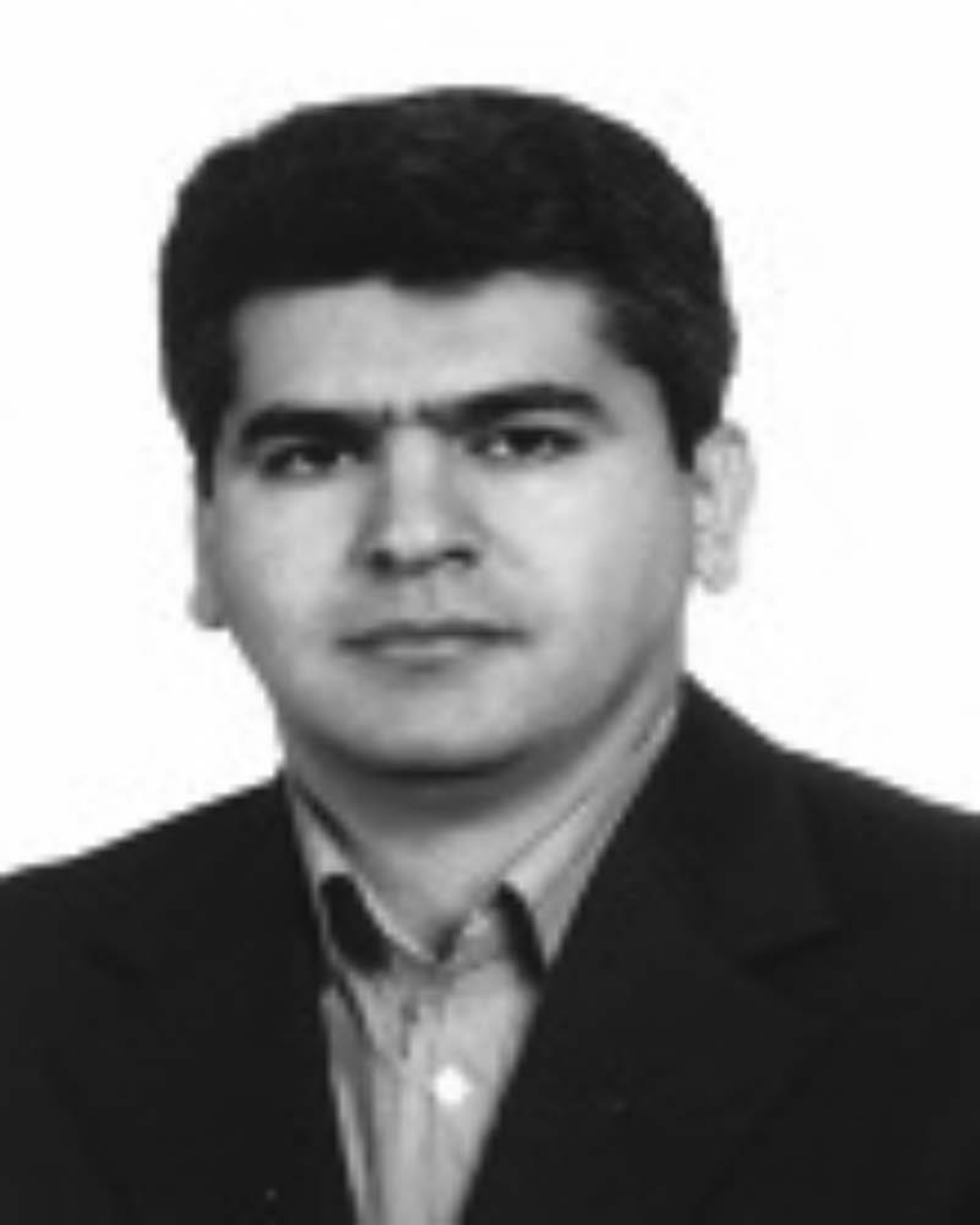}}]{Abbas Seifi} is a professor of Industrial Engineering and Management Systems at Amirkabir University of Technology in Iran. He did his BASc and MASc in Industrial Engineering at Sharif University of Technology. He received his PhD in Systems Design Engineering from the University of Waterloo in Canada and worked there as a postdoctoral research associate for over 2 years. His teaching and research interests include optimisation and simulation of various operational research problems, data driven optimisation, data science, machine learning and system dynamics.
\end{IEEEbiography}

\vspace*{-2\baselineskip}
\begin{IEEEbiography}[{\includegraphics[width=1in,height=1.25in,clip,keepaspectratio]{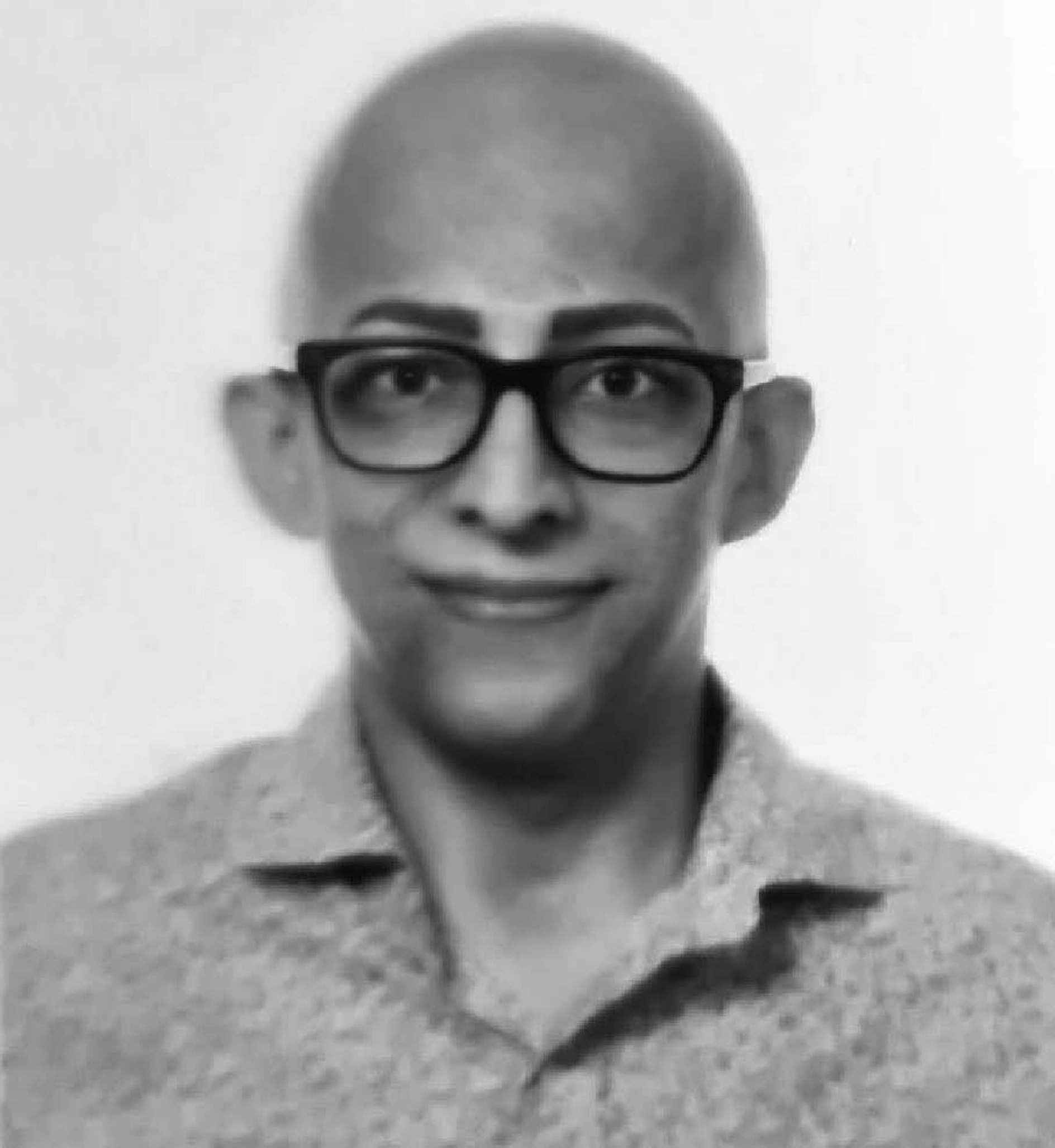}}]{Mohammad Mahdi Dehshibi}
received his PhD in Computer Science in 2017 from IAU, Iran. He is currently a research scientist at Universitat Oberta de Catalunya, Spain. He was also a visiting researcher at Unconventional Computing Lab, UWE, Bristol, UK. He has contributed to more than 60 papers published in scientific journals and international conferences. His research interests include Affective Computing, Medical data processing, Deep Learning, and Unconventional Computing.
\end{IEEEbiography}

\vfill

\end{document}